\documentclass[article]{jss}

\usepackage{orcidlink,thumbpdf,lmodern}

\usepackage{framed}
\usepackage{float}
\usepackage{bm}
\usepackage{multirow}
\usepackage{amsmath}
\usepackage{graphicx}
\usepackage{amsfonts}
\usepackage{amssymb}
\usepackage{enumerate}
\usepackage{float}
\usepackage{xcolor}
\usepackage{booktabs}
\usepackage{bbm}
\usepackage{listings}

\author{Lucas M. F. Silva~\orcidlink{0009-0000-7452-3486}\\Laborat\'orio de Matem\'atica Aplicada \\ Departamento de M\'etodos Estat\'isticos\\  Instituto de Matem\'atica \\
Universidade Federal do Rio de Janeiro 
   \And Luiz F. V. Figueiredo~\orcidlink{0009-0007-9002-1997}\\ Laborat\'orio de Matem\'atica Aplicada \\  Departamento de M\'etodos Estat\'isticos\\ Instituto de Matem\'atica \\
   Universidade Federal do Rio de Janeiro 
  \AND  Viviana G. R. Lobo~\orcidlink{0000-0002-4076-8327} \\ Laborat\'orio de Matem\'atica Aplicada\\ Departamento de M\'etodos Estat\'isticos\\ Instituto de Matem\'atica \\Universidade Federal do Rio de Janeiro 
    \And Thaís C. O. Fonseca~\orcidlink{0000-0002-4943-3259} \\ Laborat\'orio de Matem\'atica Aplicada \\ Departamento de M\'etodos Estat\'isticos\\ Instituto de Matem\'atica\\Universidade Federal do Rio de Janeiro 
   \AND Mariane B. Alves~\orcidlink{0000-0002-2489-9300}\\ Laborat\'orio de Matem\'atica Aplicada \\ Departamento de M\'etodos Estat\'isticos\\ Instituto de Matem\'atica \\Universidade Federal do Rio de Janeiro }

\Plainauthor{Lucas M. F. Silva , Luiz F. V. Figueiredo, Viviana G. R. Lobo, Tha\'is C. O. Fonseca, Mariane, B. Alves  }

\title{BayesMortalityPlus: A package in \proglang{R} for Bayesian mortality modelling}
\Plaintitle{BayesMortalityPlus: A package in R for Bayesian graduation of mortality modelling}
\Shorttitle{BayesMortalityPlus: A package in \proglang{R} for Bayesian graduation of mortality modelling}

\Abstract{
The \textbf{BayesMortalityPlus} package provides a framework for modelling and predicting mortality data. The package includes tools for
the construction of life tables based on Heligman-Pollard laws, and also on dynamic linear smoothers. Flexibility is available in terms of modelling so that the response variable may be modeled as Poisson, Binomial or Gaussian. If temporal data is available, the package provides a Bayesian implementation for the well-known Lee-Carter model that allows for estimation, projection of mortality over time, and assessment of uncertainty of any linear or nonlinear function of parameters such as life expectancy. Illustrations are considered to show the capability of the proposed package to model mortality data.
}

\Plainkeywords{Bayesian mortality graduation, Heligman-Pollard model, Dynamic linear model, Bayesian Lee-Carter model, R}

\Address{
 Viviana G. R. Lobo\\
  Departmento de Métodos Estatísticos\\
  \emph{and}\\
  Laboratório de Matemática Aplicada\\
  Instituto de Matemática\\
 Universidade Federal do Rio de Janeiro\\
  Av. Athos da Silveira Ramos, Centro de Tecnologia, Bloco C, CEP 21941-909.\\
  E-mail: \email{viviana@dme.ufrj.br}\\
  URL: \url{https://sites.google.com/a/dme.ufrj.br/viviana/}
}

\begin{document}






\section[Introduction]{Introduction} \label{sec:intro}




Models used to characterize mortality data through the Bayesian paradigm have become more popular and called the attention of actuaries, statisticians, and other researchers in recent years. In the actuarial context, it is essential to understand the mortality behaviour so that smoothed death probabilities over ages can be used for pricing life insurance and annuities. From a demographic point of view, this is an essential tool for understanding the changes of patterns in a population. Thus, applying mathematical formulations, such as mortality laws, smoothing models, and improvement techniques is useful to understand the mortality curves of populations or portfolios. Statistical methodologies that consider Bayesian graduation have been more attractive as it allows for the incorporation of prior knowledge through the prior distributions. 
Besides, graduation is particularly important at advanced ages, for which exposure numbers are small and data are sparse, see \cite{Forster2018}.

 Mortality graduation models have become more sophisticated over time. \cite{Kimeldorf1967} propose the use of mortality smoothing and the constructions of life tables via Bayesian graduation, and \cite{Carlin1992} proposes the use of Markov chain Monte Carlo (MCMC) techniques to fit mortality curves. \cite{dellaportas2001bayesian} suggest estimating the Heligman–Pollard (HP) laws proposed by \cite{heligman1980age} using a non-linear logistic and Log-Normal model that accounts for uncertainty in model parameters. \cite{Njenga2011} use the Bayesian vector auto-regressive (BVAR) model for the parameters of the HP model by considering temporal evolution of parameters in the HP function. \cite{Forster2018} and \cite{HiltonForster2019} provide a methodology for mortality estimation based on generalized additive models \citep[GAMs - see][]{Wood2006} at the youngest ages and use a simpler parametric model at older ages that depend on mortality laws well-established in the literature. Packages and functions for fitting mortality curves have been available in \proglang{R} environment \citep{RCoreTeam} for several years, through the Comprehensive R Archive Network (CRAN). The \pkg{MortalityLaws} package exploits optimization methods for fitting a wide range of point estimates for mortality laws \citep{mortlaws}. Recently, this package was removed from CRAN repository and old versions can be accessed at \url{https://cran.r-project.org/src/contrib/Archive/MortalityLaws/}. The \pkg{demography} package developed by \cite{demography} provides functions for demographic analysis, such as life table calculations, fertility rates, and functional data analysis of mortality rates. In the context of Bayesian computation, R packages have been proposed such as the \pkg{HPbayes} package that provides the eight parameters of the Heligman-Pollard mortality model using a Bayesian Melding procedure with importance sampling \citep{HPbayes}. However, the \pkg{HPbayes} package is no longer available in the \proglang{R} CRAN repository. Formerly available versions can be obtained from \url{https://cran.r-project.org/src/contrib/Archive/HPbayes/}. 

The Heligman-Pollard law considers a specific mathematical function to model mortality rates, as a mixture of infant, young adult, and adult survival functions. However, other flexible smoothing approaches could be considered to model mortality, such as splines techniques \citep{Eilers2004,Carmada2016,Carmada2019,Forster2021}. In this context, \cite{mortsmooth} proposes a package in \proglang{R} called \pkg{MortalitySmooth}, that provides a framework for smoothing count data assumed to be Poisson-distributed in both one- and two-dimensional settings through P-splines.  In addition to the proposal of a function in \pkg{BayesMortalityPlus} to model mortality curves via the Heligman-Pollard law, in this article, we propose a smoother based on dynamic linear models (DLM) \citep{West97} that is flexible as splines and has an interpretable parameter for controlling smoothness in the mortality graduation across ages. 
Dynamic linear models are a large class of models with time-varying parameters, useful for modelling time series data. Basically, the proposal is to consider the age of the individuals as an indexer term rather than the evolution over time. More details about the smoothing proposed model are described in Section \ref{sec3}.

Although the Heligman-Pollard model is well-known for forecasting future mortality rates, there are other models that can be taken into account. Among  several methods, the Lee-Carter model \citep{Lee1992} is a stochastic demographic model that considers temporal dependence in the data, whereas the Heligman-Pollard is a parameterization function for cross-section data.  
This was a pioneer work in the mortality modelling of a single population over time. The method is based on a factor model with a latent factor varying across time (a state parameter). Several extensions have been proposed to the Lee-Carter model. \cite{LiLee04} present an extension of the Lee-Carter model which allows for mortality prediction when the time series is observed in unequal intervals of time. The paper discusses the effects of parameter estimation and prediction uncertainty when the data is limited. The package \pkg{demography},  previously mentioned, implements the original Lee-Carter model and other variants presented in {\cite{LeeMiller2001}, \cite{Booth2002}, and \cite{Hydman2007}}. The package \pkg{StMoMo} developed by \cite{stmomo} fits the Lee-Carter model amongst a handful of other mortality models via generalized non-linear models, using the existent \pkg{gnm} \proglang{R} package \citep{gnm}. From a Bayesian point of view, several papers have dealt with mortality modelling, such as \cite{Czado2005} and \cite{Pedroza2006}. In this way, the package \pkg{StanMoMo} \citep{stanmomo} models a variety of popular stochastic models with the help of Stan software via \pkg{rstan} \citep{rstan}. However, it does require some degree of knowledge of Stan tools to perform mortality graduation.

 Although several packages to study mortality data are available, there are some issues that our proposed package \pkg{BayesMortalityPlus} seeks to solve.  Firstly, our package provides a user-friendly interface, as well as instructions and simple examples for running each function available for mortality modelling and prediction. 
 The package provides examples from the Human Mortality Database \citep{hmd}, and allows the user to supply external data if desired.
 Moreover, we perform the full Bayesian inference procedure for several smoothing and prediction models: for the HP laws following the specifications described in \cite{dellaportas2001bayesian}, the Lee-Carter model, as described by \cite{Pedroza2006} and the Dynamic Linear Model via forward filtering backward sampling (FFBS) recursions with Gibbs sampling, described in \cite{Carter1994} and \cite{Sylvia1994}, and proposed here for mortality graduation. This approach makes it possible to simplify the modelling process for the user, as well as provide estimation and credible intervals of parameters and nonlinear functions of model parameters with correct uncertainty measurement, such as probabilities of death, life expectancy, and an easy visualization through graphic tools. We also provide a specific function for closing life tables that is not available in any of the packages mentioned previously. It is based on the article by \cite{Forster2018} and aims to obtain a more robust fit for advanced adult ages, for which exposures are usually reduced.
 
In brief, we review the statistical framework underlying the \textbf{BayesMortalityPlus} package and show its ability to model mortality rates via the Heligman-Pollard laws as suggested by \cite{dellaportas2001bayesian}, the Dynamic linear model, and the Lee-Carter model via the Bayesian framework. The inference procedure considers the use of Markov chain Monte Carlo techniques to estimate the mortality curve and to perform prediction. The package was coded in \proglang{R} and is available on CRAN 
\url{https://cran.r-project.org/package=BayesMortalityPlus}. Version {0.1.0} has been considered for this article.

The remaining text is organized as follows. Section \ref{sec2} presents the Bayesian Heligman-Pollard model, beginning with a brief review of the original Heligman-Pollard law and its properties. The Bayesian inferential and computational procedures based on Monte Carlo Markov chain techniques are addressed. Section \ref{sec3} presents the Bayesian graduation via Dynamic Linear Model where we propose to model the mortality rates with age-indexes replacing the usual temporal dimension. Section \ref{sec4} provides tools to the Bayesian graduation through \pkg{BayesMortalityPlus} package. Section \ref{sec4.1} shows an illustrative example modelling the mortality curve via the HP model. The functions available in the package are presented and applied to purposes such as computing life expectations, closing life tables using different methods, and producing plots. In Section \ref{sec4.2}, the same example is considered using the Dynamical linear models. Section \ref{sec5} introduces the Bayesian Lee-Carter model and some functions supplied in the package are employed. Section \ref{sec6} concludes with a final discussion and some remarks.

\section[Bayesian graduation via Heligman-Pollard model]{Bayesian graduation via Heligman-Pollard model}\label{sec2}

The methods considered in the \textbf{BayesMortalityPlus} package assume that the process of mortality tables graduation is based on a probabilistic approach that allows the computation of point estimates for mortality rates and life expectations, as well as the measurement of the associated uncertainty. We consider the data $(x, E_x, D_x, m_x)$ in a fixed period of time, where $x$ denotes the age of the individuals and assumes an integer value, $D_x$ denotes the number of deaths at age $x$ and $E_x$ denotes the total exposure of individuals aged $x$. The central mortality rate is defined as $ m_x = D_x / E_x $ \citep{bowers1986actuarial} and the probability of death as $ q_x = 1-e ^ {- m_x} $.

\begin{figure}[!ht]
    \centering
    \includegraphics[width = 0.9\textwidth]{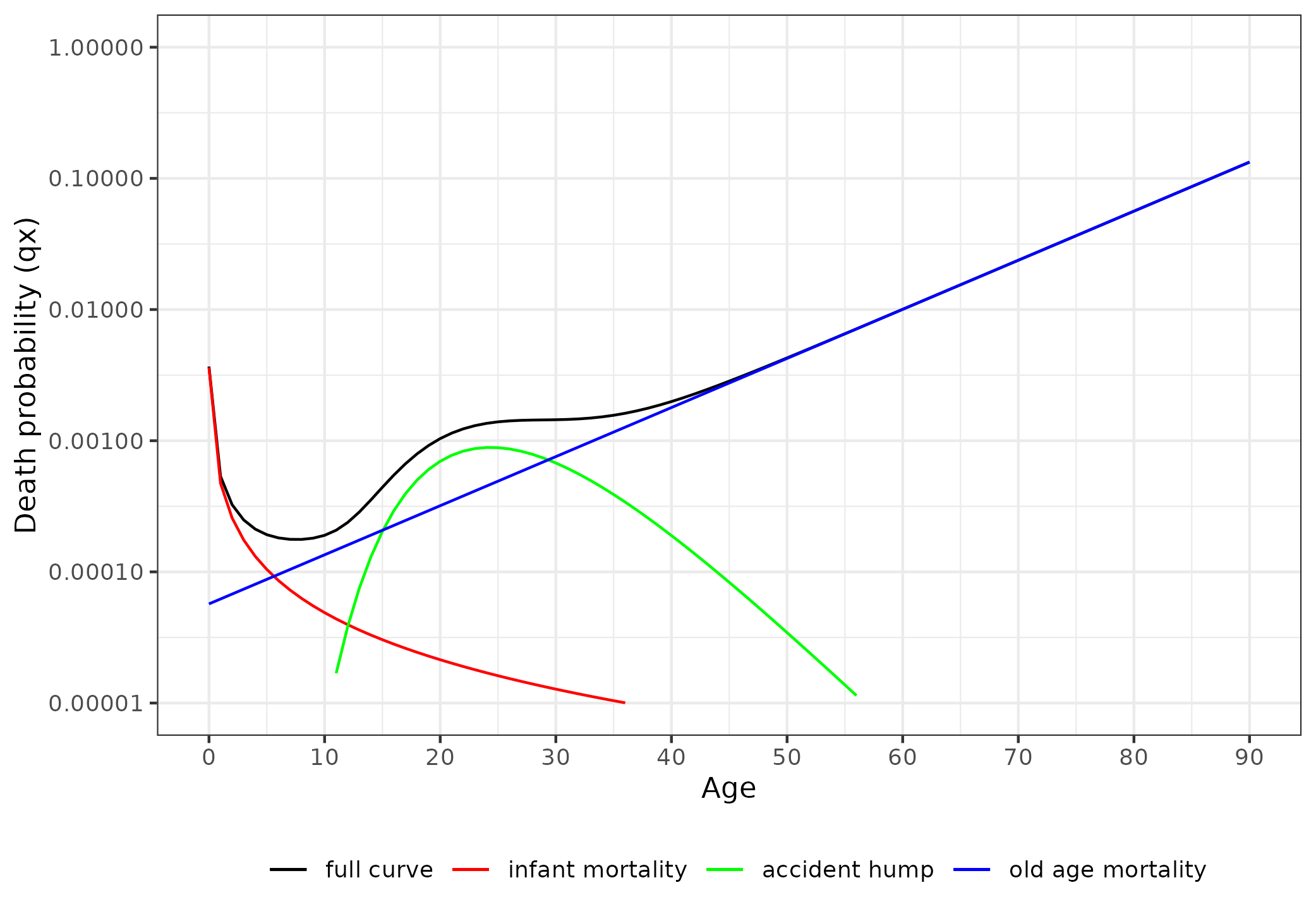}
    \caption{An illustration of the Heligman-Pollard curve: progress with the age of the logarithm of the probability of dying and of the logarithms of its three components. The formula were evaluated approximately at $(A, B, C, D, E, F, G, H) = (5.04 \times 10^{-4},~~7.49 \times 10^{-2},~~1.16 \times 10^{-1},~~8.89 \times 10^{-4},~~6.28,~~24.35,~~5.7 \times 10^{-5},~~1.09)$. }
    \label{fig:hpcurve}
\end{figure}

An usual approach considered for mortality graduation uses the well-known Heligman-Pollard law \citep{heligman1980age}.  The HP model is a parametric function that captures the main characteristics of mortality tables, specified in terms of parameters that aim to have a demographic interpretation of the ages' effect on the mortality rates. It is written as
\begin{equation}\label{sec2:eq1}
\frac{q_x}{1- q_x} = A^{(x+B)^{C}} + D \thinspace exp\left[ - E \left\{log \left(\frac{x}{F} \right) \right\}^2\right] + G H^x.
\end{equation}
Equation (\ref{sec2:eq1}) provides a mathematical formulation that takes into account three terms, each representing a mortality component over the age domain as illustrated in Figure \ref{fig:hpcurve}. 
The first term reflects the fall in mortality during the early childhood years through a rapidly declining exponential curve. The second term, similar to the Log-Normal curve, reflects accident mortality for males and accident plus maternal mortality for the female population, being called the \textit{accident hump}. The last term reflects the near geometric rise in mortality experienced in advanced ages, through the Gompertz exponential formula as described in \cite{heligman1980age}. Table \ref{tab:HP_xerms} summarises the interpretation of each term of the HP model. 
\begin{table}[t!]
\centering
\begin{tabular}{ p{5cm} p{4cm}  p{5cm} }
\hline
Term & Interpretation & Parameters \\
\hline
\\
\multirow{6}{*}{$A^{(x+B)^{C}}$} & \multirow{6}{*}{Infant mortality} & \textbf{$A$} measures the level of the mortality.
\newline \textbf{$B$} is an age displacement for the mortality of an infant (age 1). 
\newline \textbf{$C$} measures the decline of the mortality rate throughout childhood.
\newline The parametric domain of these three parameters lies on the interval (0, 1)\\
\\
\multirow{3}{*}{$D \thinspace exp\left[ - E \left\{log \left(\frac{x}{F} \right) \right\}^2\right]$} & \multirow{3}{*}{"Accident hump"} & \textbf{$D$} represents the severity, \textbf{$E$} represents the spread and F the location of the "accident hump". These three parameters have the following domains: $D \in (0,1)$, $E \in (0, \infty)$ and $F \in (15, 110)$.\\
\\
\multirow{3}{*}{$G H^x$} & \multirow{3}{*}{Advanced age mortality} & \textbf{$G$} represents the base level of senescent mortality while \textbf{$H$} reflects the rate of increase of that mortality. Their respective domains are: $G \in (0,1)$ and $ H \in (0, \infty) $.\\
\\
\hline
\end{tabular}
\caption{Description of the parameters of the Heligman Pollard model.}
\label{tab:HP_xerms}
\end{table}

To estimate the parameters in equation (\ref{sec2:eq1}) several methods have been proposed. The first method suggested by \cite{heligman1980age} considers weighted least squares with weights $w_x = 1 / \hat{q}_x^2$. This proposal could be problematic due to the over-parameterization of the model and numerical instabilities. \cite{HPbayes} considers the  Bayesian Melding with Incremental Mixture Importance Sampling techniques implemented in the \textbf{HPbayes} package. \cite{dellaportas2001bayesian} suggest Bayesian inference using the Markov Chain Monte Carlo method to estimate the parameters. For more details on MCMC algorithms see \cite{gamerman}.


Following the proposal based on \cite{dellaportas2001bayesian}, we assume that the death odds are modelled through the Log-Normal distribution and that all individuals of the same age die independently with the same probability and a constant parameter of variation $\sigma^2$ for all ages. Therefore, the model can be written as
\begin{equation}\label{sec2:eq2}
log \left(\frac{q_x}{1- q_x}\right) = log \left( A^{(x+B)^{C}} + D \thinspace exp\left[ - E \left\{log \left(\frac{x}{F} \right) \right\}^2\right] + G H^x \right) + \varepsilon_x,
\end{equation}
where $\varepsilon_x \sim N(0, \sigma^2)$ are independent for all age $x$. Equation (\ref{sec2:eq2}) can be rewritten in a general form as $log(y_x) = log(f_x) + \varepsilon_x$, with $f_x$ being a parametric function. Thus $E(y_x)=f_x\exp(\sigma^2/2)$ and $Var(y_x)=[\exp(\sigma^2)-1]\exp(\sigma^2)f_x^2$. Here the Markov chain Monte Carlo techniques require the updating of parameters that depend on the function $f_x$ and the parameter $\sigma^2$. See \cite{dellaportas2001bayesian} for a more detailed discussion.

Although the proposal in \cite{dellaportas2001bayesian} considers modelling the odds via a log-normal distribution, several papers make other probabilistic assumptions about the mortality law (\cite{Czado2005}, \cite{Renshaw1996}, \cite{Li2013}). In particular, we are interested in allowing the exposure to be related to the model uncertainty since lower exposure is usually associated with higher variability in the data. Therefore we consider modelling the mortality via Poisson and Binomial models as suggested by \cite{dellaportas2001bayesian}.

The Binomial model assumes that $D_x$, which denotes the death count at age $x$, follows a Binomial distribution with the size parameter being the exposure in age $x$, that is, $E_x$, with death probability at age $x$ given by $q_x$. On the other hand, the Poisson model considers that $D_x$ represents the death counts for the age $x$ following a Poisson distribution with rate $E_x \times q_x$ for each age. In this case, the exposure is an offset. For these two sampling distributions, we will consider an alternative representation for the HP curve by adding an extra parameter as follows
\begin{equation}\label{sec2:eq3}
q_x = A^{(x+B)^{C}} + D \thinspace exp\left[ - E \left\{log \left(\frac{x}{F} \right) \right\}^2\right] + \frac{G H^x}{1 + KG H^x},
\end{equation}
where the parameter $K$ is considered in order to allow for changes in the concavity of the curve at its final portion, resulting in a more flexible approach for capturing mortality trends at advanced ages. These alternative formulations have the advantage that the uncertainty relative to the mortality data changes according to the exposure at each age.

In \pkg{BayesMortalityPlus} package, the user can estimate the parameters of the HP curve through the function \code{hp} for the Log-Normal, Binomial, and Poisson models. \pkg{BayesMortalityPlus} can be installed with the code:
\begin{CodeChunk}
\begin{CodeInput}
R> install.packages("BayesMortalityPlus")
\end{CodeInput}
\end{CodeChunk}
The package is loaded within \proglang{R} as follows:
\begin{CodeChunk}
\begin{CodeInput}
R> library("BayesMortalityPlus")
\end{CodeInput}
\end{CodeChunk}
The function reproduces the inference procedure presented by \cite{dellaportas2001bayesian} as follows
\begin{Code}
hp(x, Ex, Dx, model = c("binomial", "lognormal", "poisson"),
    M = 50000, bn = round(M/5), thin = 10, m = rep(NA, 8), 
    v = rep(NA, 8), inits = NULL, K = NULL, sigma2 = NULL,
    prop.control = NULL, reduced_model = FALSE)
\end{Code}
\begin{itemize} 
\item The arguments \code{x}, \code{Ex}, and \code{Dx} represent the vector of the ages,  exposures by age, and deaths by age, respectively.
\item The argument \code{model} defines the mortality model chosen by the user.  Setting \code{model = "poisson"} assumes that deaths follow the Poisson distribution, setting \code{model = "binomial"} assumes that deaths follow the Binomial distribution, and setting \code{model = "lognormal"} assumes that the odds follow the Log-Normal distribution.
\item The arguments \code{m} and \code{v} can be used to specify means and variances, respectively, for the prior distributions of each parameter, with \code{inits} specifying the initial values for the parameters in the algorithm. The \code{K} argument specifies the extra parameter $K$ for the Binomial and the Poisson models, while \code{sigma2} is responsible for the initial value for the variance estimated for the Log-Normal distribution. Also, the argument \code{prop.control} tunes the acceptance rate of the MCMC algorithm, for which \code{M} iterations are assumed, with burn-in period  \code{nb} and thinning given by the argument  \code{thin}. Details on the specification of the MCMC algorithm can be seen in \citet{gamerman}.
\item The argument \code{reduced_model} allows the user to fit a truncated version of the HP curve, which will be discussed in Section \ref{sec4.1}
\end{itemize}
The package makes the posterior distribution samples  available for the user to make inferences about any transformations of the parameters. Therefore, it is simple to obtain the probability of death $q_x$ for any age $x$. Furthermore, the user is able to compute predictive intervals for $q_x$ and survival probabilities $(p_x = 1 - q_x )$, which can be used to quantify the life expectancy for any required age. 

For the Binomial and the Poisson models,  the HP formula provides estimates for the central mortality rate $m_x$. Then, under the assumption of uniform distribution for the deaths over an age interval $x$, we can compute the death probability $q_x$ at age $x$ through the usual relation $q_x = 1 - exp(-m_x)$. For the Log-Normal model, these probabilities can be obtained through $q_x = \mu_x/(1 - \mu_x)$, where $\mu_x$ denotes the HP curve at age $x$. Finally, we obtain the point estimation for the death probabilities through the posterior median distribution of $q_x$. 

%
%

 %
%

\section[Bayesian graduation via Dynamic Linear Model]{Bayesian graduation via Dynamic Linear Model}
\label{sec3}
Dynamic Linear Models (DLM) \citep{West97} are usually applied in time series analysis, in order to address intrinsically auto-correlated observations gathered through times $t=1,2,\ldots$. In this work, we adopt a particular specification of the DLM class to produce graduated mortality tables, formally recognising the association between mortality rates for neighbouring ages (or age groups) and imposing smoothness of the estimated mortality curve in the transition between ages. The use of DLMs in mortality studies has been common practice in modelling temporal dependence and has been used for mortality predictions. For instance, the well-known Lee-Carter model \citep{Lee1992} considers a dynamical model to estimate temporal improvement for each age and can be used for predicting mortality in future years.  
Other proposals are \cite{LiLee04} and \citet{MigonNeves2007}. 
The point to be highlighted here is that  in our approach  dynamic components are indexed by ages and not by time periods, aiming to obtain smooth non-linear graduated curves through ages, as well as to address the autocorrelation among ages. Let $D_x$ and $E_x$, respectively, denote the death counts and exposure at age $x$ and define $Y_x=\log \left(\frac{D_x}{E_x}\right)$. We consider a second-order polynomial DLM, as follows: 
\begin{eqnarray}
Y_x &=& \mu_x + v_x,~~~~v_x \sim N(0,V) \label{eq.obs}\\
\mu_x &=& \mu_{x-1} + \beta_{x-1} + w_{x,1},~~~~w_{x,1} \sim N(0,\sigma^2_{\mu,x}) \label{eq.evol.mu}\\
\beta_x &=& \beta_{x-1} + w_{x,2},~~~~w_{x,2} \sim N(0,\sigma^2_{\beta,x}), \label{eq.evol.beta}
\end{eqnarray}
with random errors $v_x$, $w_{x,1}$ and $w_{x,2}$ assumed mutually and sequentially independent.
The state $\mu_x$ denotes the dynamic level of the log mortality, with stochastic evolution guided by equation (\ref{eq.evol.mu}), and $\beta_x$ controls the level variation between consecutive ages (local slope of the mortality curve), allowing for different gradients through ages since $\beta_x$ evolves according to the random walk described in equation (\ref{eq.evol.beta}).  The smoothness of the graduated mortality curves strongly depends on the magnitude of the evolutional errors' variances, $\sigma^2_{\mu,x}$ and $\sigma^2_{\beta,x}$, which are specified via discounting strategies, as discussed in \citet[][Chapter 6]{West97}.  

The model may be rewritten, in the general DLM form, as 
\begin{eqnarray*}
Y_x &=& \bm F_x\bm \theta_x +  v_x,~~~~v_x \sim N(0,V)\\
\bm \theta_x &=& \bm G_x\bm  \theta_{x-1} +\bm w_x,~~~~\bm w_{x} \sim N_2(\bm0,\bm W_x),
\end{eqnarray*}
where $N_2(\cdot,\cdot)$ denotes a bivariate Gaussian density and
\begin{equation*}
\bm \theta_x = 
\begin{bmatrix}
\mu_x \\ \beta_x 
\end{bmatrix},
~~~~\bm G_x = 
\begin{bmatrix}
1 & 1 \\ 0 & 1
\end{bmatrix},
~~~~\bm W_x = 
\begin{bmatrix}
\sigma^2_{\mu,x} & 0 \\ 0 & \sigma^2_{\beta,x}
\end{bmatrix},
~~~~\bm F_x = 
\begin{bmatrix}
1 & 0
\end{bmatrix}, \quad x=1,2,\ldots.
\end{equation*}

Details on general forms of the DLMs and specially on the particular case of polynomial trend models, adopted here, are found in \citet[][Chapter 7]{West97} and \citet[][Section 3.2.1]{petris2009dynamic}. A first-order polynomial model, that is, a model with only a dynamic $\mu_x$ could be able to capture several dynamic mortality patterns over the ages, but the resulting point predictive function for the following $h$ ages would be a constant function of $h$. The use of the additional parameter $ \beta_x$ results in a DLM that generates a predictive curve for $h$ future ages given by a non-zero slope straight line, thus capturing the increasing mortality risk for advanced ages. The concern with the form of the predictive function associated with the adopted model is justified by the fact that, for advanced ages, it is usual that the databases present a shortage of exposure. Therefore the mortality tables are typically adjusted using information up to a certain age $x_*$ and from that point on, extrapolations are necessary. In \pkg{BayesMortalityPlus}, the predictive function of the second-order polynomial DLM is used in the extrapolation process of the mortality curve.

The user can estimate the parameters of the DLM through the function \code{dlm}. The function implements the inference procedure based on \cite{West97} via Gibbs sampling for state space models presented by \cite{Carter1994} and \cite{Sylvia1994} as follows
\begin{Code}
dlm(y, Ft = matrix(c(1,0), nrow = 1), Gt = matrix(c(1,0,1,1), 2),
  delta = 0.85, prior = list(m0 = rep(0, nrow(Gt)), C0 = diag(100, nrow(Gt))),
  prior.sig2 = list(a = 0.01, b = 0.01), M = 5000, bn = 3000, thin = 1,
  ages = 0:(length(y)-1))
\end{Code}
\begin{itemize} 
\item \code{y} represents the vector of log mortality rates.
\item The arguments \code{Ft}, \code{Gt} and \code{delta} represent the structural elements for the  specification of the observational and system equations, and the discount factor (default =0.85) for the smoothing, respectively. 
\item The arguments \code{prior} and \code{prior.sig2} can be used to specify prior information, both as a \code{list} object. Argument \code{prior} receives the prior mean vector and covariance matrix and \code{prior.sig2} receives the prior parameters of the Inverse Gamma distribution for the estimated variance of the process.
\item The argument \code{ages} allows the user to define the vector of ages associated with \code{y} in case the age interval does not equal the default graduation \code{0:(length(y)-1)}.
\end{itemize}

Samples from the posterior distribution are available for inference about quantities of interest such as the probabilities of death $q_x$, the associated predictive intervals, and the survival probabilities $(p_x = 1 - q_x)$.

For the DLM approach, the model fit provides samples of the posterior distribution of $\mu_x$, from which  a point estimate of the log mortality rate $y_x$ can be computed. Since posterior samples are available, the death probability $q_x$ at age $x$ can be obtained through the relation $q_x = exp(y_x)$. For instance, the posterior median of $q_x$ can be used as the point estimation for the death probabilities. 
%
%
 %
%

\newpage

\section[Static graduation with BayesMortalityPlus]{Static graduation with BayesMortalityPlus}
\label{sec4}
In this Section, we present the functions available in  \pkg{BayesMortalityPlus} that can be used in the construction of life tables based on the Heligman-Pollard law and the Dynamic Linear models, respectively, as described in Sections \ref{sec2} and \ref{sec3}. The main functions for smoothing are \code{hp} and \code{dlm}. We also explore the posterior summaries and methodologies for extrapolation. Data from the United States and Portugal, which are extracted from the Human Mortality Database \citep{hmd}, are contained in the object \code{data}, stratified by  sex (as well as total population). 

In the following, we present the Bayesian graduation by selecting the total population from the United States over the past forty years. We estimate the mortality curves for the specific years 1980, 1990, 2000, 2010, 2019: 

\begin{CodeChunk}
\begin{CodeInput}
R> library(BayesMortalityPlus)
R> data(USA)
\end{CodeInput}
\end{CodeChunk}
We load the \pkg{dplyr} package to extract information from the database in a simple way through the command \code{filter} \citep[see more details in][]{dplyr}. Notice that other ways to manipulate the data could be applied. In this example, the vector \code{[1:81]} means that the ages $x=0, \ldots, 80$ are selected, so that the exposures ($E_x$) and death counts ($D_x$) for the years considered in the study are specified and filtered up to 80 years old for model fitting. 
\begin{CodeChunk}
\begin{CodeInput}
R> ex_1980 <- dplyr::filter(USA, Year == 1980)$Ex.Total[1:81]
R> dx_1980 <- dplyr::filter(USA, Year == 1980)$Dx.Total[1:81]
R> ex_1990 <- dplyr::filter(USA, Year == 1990)$Ex.Total[1:81]
R> dx_1990 <- dplyr::filter(USA, Year == 1990)$Dx.Total[1:81]
R> ex_2000 <- dplyr::filter(USA, Year == 2000)$Ex.Total[1:81]
R> dx_2000 <- dplyr::filter(USA, Year == 2000)$Dx.Total[1:81]
R> ex_2010 <- dplyr::filter(USA, Year == 2010)$Ex.Total[1:81]
R> dx_2010 <- dplyr::filter(USA, Year == 2010)$Dx.Total[1:81]
R> ex_2019 <- dplyr::filter(USA, Year == 2019)$Ex.Total[1:81]
R> dx_2019 <- dplyr::filter(USA, Year == 2019)$Dx.Total[1:81]
\end{CodeInput}
\end{CodeChunk}
\begin{figure}[t!]
    \centering
    \includegraphics[width = 0.9\textwidth]{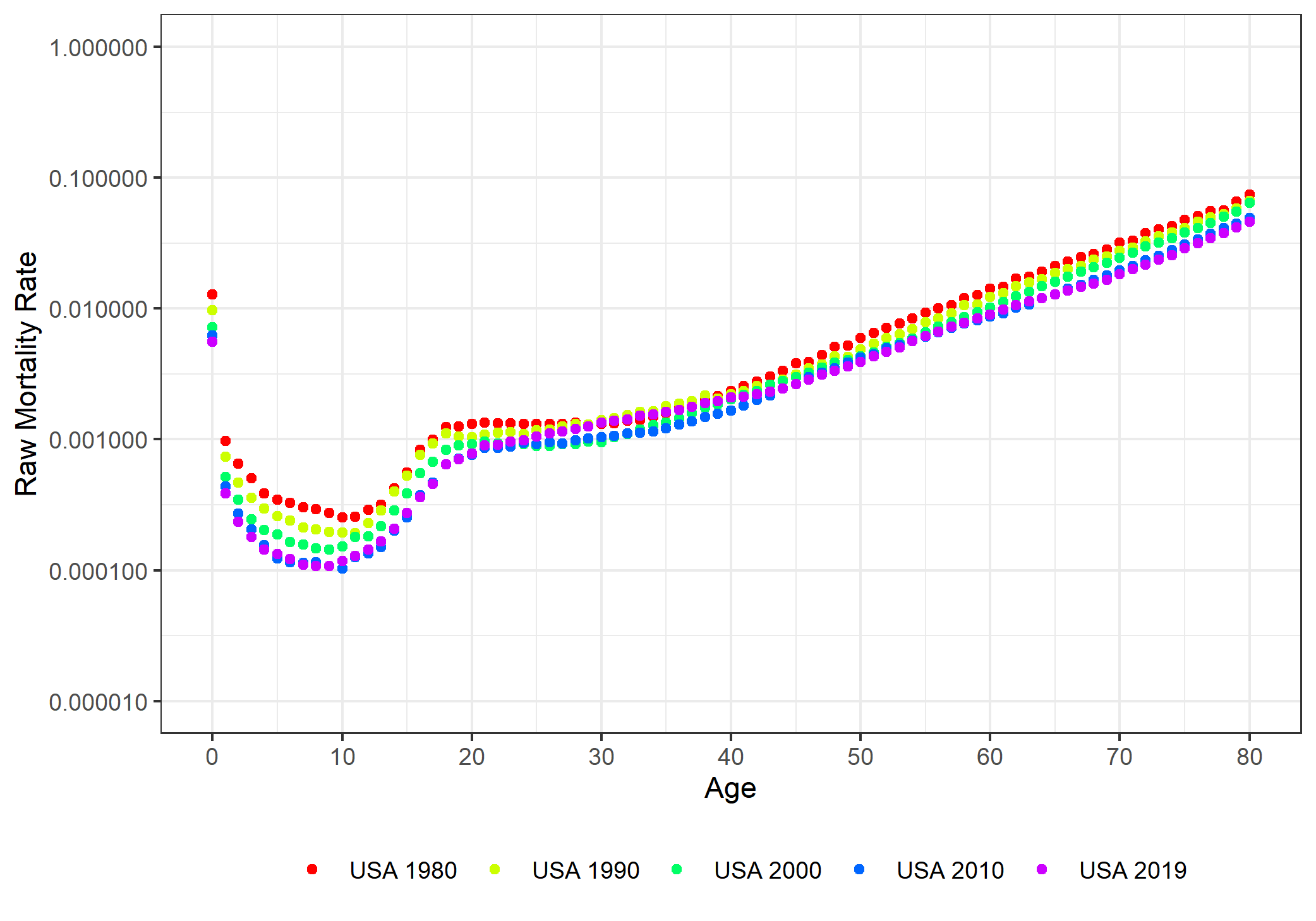}
    \caption{Raw mortality rates in log-scale. The United States, total population, ages 0-80 and years 1980, 1990, 2000, 2010 and 2019.}
    \label{fig:raw_xables}
\end{figure}
Figure \ref{fig:raw_xables} illustrates the raw mortality rates ($q_x = D_x/E_x$) over years via command \code{ggplot} available on \pkg{gglplot} package with the following code:
\begin{CodeChunk}
\begin{CodeInput}
R> qx_1980 <- dx_1980/ex_1980
R> qx_1990 <- dx_1990/ex_1990
R> qx_2000 <- dx_2000/ex_2000
R> qx_2010 <- dx_2010/ex_2010
R> qx_2019 <- dx_2019/ex_2019
R> data = data.frame(idade = 0:80, qx_1980 = qx_1980,
                  qx_1990 = qx_1990, qx_2000 = qx_2000,
                  qx_2010 = qx_2010, qx_2019 = qx_2019)
R> data = as.data.frame(data) 
R> ggplot(data) +
  scale_y_continuous(trans = "log10", breaks = 10^-seq(0,5),
                     limits = 10^-c(5,0), labels = scales::comma) +
  scale_x_continuous(breaks = seq(0, 100, by = 10)) + 
  theme_bw() + theme(legend.position = "bottom") +
  labs(x = "Age", y = "Raw Mortality Rate", title = NULL) +
  geom_point(aes(x = idade, y = qx_1980, col = "1")) +
  geom_point(aes(x = idade, y = qx_1990, col = '2')) +
  geom_point(aes(x = idade, y = qx_2000, col = "3")) +
  geom_point(aes(x = idade, y = qx_2010, col = "4")) +
  geom_point(aes(x = idade, y = qx_2019, col = "5")) +
  scale_color_manual(name = NULL, values = c(rainbow(5)),
                     label = c("USA 1980", "USA 1990", "USA 2000",
                               "USA 2010","USA 2019"))
\end{CodeInput}
\end{CodeChunk}

\subsection[Heligman-Pollard model]{Heligman-Pollard model} \label{sec4.1}

The function \code{hp} returns an object of class \code{"HP"}, which is an HP curve fit to the input data settled by the user. In this illustration, we consider vague or non-informative prior distributions. Notice that the user could provide their own prior information if desired. The MCMC scheme is the default one. The HP model under a Log-Normal setting for the respective years can be defined using the following code: 
\begin{CodeChunk}
\begin{CodeInput}
R> fit_1980 <- hp(0:80, ex_1980, dx_1980, model = "lognormal")
Simulating [===================================] 100% in  1m
R> fit_1990 <- hp(0:80, ex_1990, dx_1990, model = "lognormal")
Simulating [===================================] 100% in  1m
R> fit_2000 <- hp(0:80, ex_2000, dx_2000, model = "lognormal")
Simulating [===================================] 100% in  1m
R> fit_2010 <- hp(0:80, ex_2010, dx_2010, model = "lognormal")
Simulating [===================================] 100% in  1m
R> fit_2019 <- hp(0:80, ex_2019, dx_2019, model = "lognormal")
Simulating [===================================] 100% in  1m
\end{CodeInput}
\end{CodeChunk}
The \code{summary} function in \proglang{R} provides a summary table with the estimation of the parameters and the acceptance rate of the MCMC algorithm. As an example, the posterior summary for the 1980-year fit is available using the code:
\begin{CodeChunk}
\begin{CodeInput}
R> summary(fit_1980)
       mean       sd      2.5%     50.0%     97.5% Accept %
A  0.001027 0.000053  0.000929  0.001024  0.001134     22.3
B  0.026701 0.006329  0.016337  0.025984  0.040566     22.3
C  0.125931 0.005250  0.115776  0.125809  0.136760     22.3
D  0.000923 0.000031  0.000864  0.000923  0.000987     22.3
E 11.499650 0.697406 10.202269 11.493265 12.926974     22.3
F 21.088628 0.155102 20.791117 21.085325 21.403002     22.3
G  0.000074 0.000003  0.000069  0.000074  0.000079     22.3
H  1.090734 0.000677  1.089432  1.090735  1.092063     22.3 
\end{CodeInput}
\end{CodeChunk}
%
The \code{fitted} function can provide a summary with the point estimate of death probabilities generated by the model for specific ages:
\begin{Code}
fitted(fit, age = NULL)
\end{Code}
The argument \code{fit} is a fitted curve by the \code{hp} function via \pkg{BayesMortalityPlus} package, and the argument \code{age} represents the age interval in which the estimation of the death probabilities is desired. The default age interval is set to \code{NULL}, which means that the function will return the whole age interval fitted by the model. 
For illustration, consider setting the ages \code{0,20,40,60,80} and the year 1980.
\begin{CodeChunk}
\begin{CodeInput}
R> fitted(fit_1980, age = c(0,20,40,60,80))
  age   qx_fitted
1   0 0.012809519
2  20 0.001354225
3  40 0.002402713
4  60 0.013353342
5  80 0.071351681
\end{CodeInput}
\end{CodeChunk}
%
\begin{table}[t!]
\centering
\begin{tabular}{lccccc}
\hline
 & US 1980 & US 1990 & US 2000 & US 2010 & US 2019\\
\hline
\multirow{2}{*}{$A$} & 0.00103 & 0.00077 & 0.00054 & 0.00051 & 0.0004\\
 & \footnotesize (0.00093; 0.00113) & \footnotesize (0.00065; 0.00091) & \footnotesize (0.00048; 0.00062) & \footnotesize (0.00041; 0.00063) & \footnotesize (0.00033; 0.00048)\\
\multirow{2}{*}{$B$} & 0.0267 & 0.0381 & 0.0538 & 0.0901 & 0.0557\\
 & \footnotesize (0.0163; 0.0406) & \footnotesize (0.0171; 0.0697) & \footnotesize (0.0324; 0.0798) & \footnotesize (0.048; 0.1477) & \footnotesize (0.0251; 0.1)\\
\multirow{2}{*}{$C$} & 0.1259 & 0.1328 & 0.1432 & 0.1648 & 0.141\\
 & \footnotesize(0.1158; 0.1368) & \footnotesize(0.1157; 0.1511) & \footnotesize(0.1289; 0.1583) & \footnotesize(0.1429; 0.1895) & \footnotesize(0.1186; 0.1656)\\
\multirow{2}{*}{$D$} & 0.00092 & 0.00077 & 0.0006 & 0.00058 & 0.0008\\
 & \footnotesize(0.00086; 0.00098) & \footnotesize(0.0007; 0.00085) & \footnotesize(0.00056; 0.00065) & \footnotesize(0.00053; 0.00065) & \footnotesize(0.00073; 0.00089)\\
\multirow{2}{*}{$E$} & 11.49 & 6.43 & 11.81 & 8.79 & 4.45\\
 & \footnotesize(10.2; 12.92) & \footnotesize(5.21; 7.98) & \footnotesize(9.93; 13.78) & \footnotesize(7.1; 10.9) & \footnotesize(3.61; 5.43)\\ 
\multirow{2}{*}{$F$} & 21.08 & 22.8 & 20.9 & 23.7 & 29\\
 & \footnotesize(20.79; 21.4) & \footnotesize(21.97; 23.81) & \footnotesize(20.54; 21.36) & \footnotesize(22.96; 24.57) & \footnotesize(27.54; 30.77)\\ 
\multirow{2}{*}{$G$} & 0.00007 & 0.00006 & 0.00006 & 0.00005 & 0.00004\\
 & \footnotesize(0.00007; 0.00008) & \footnotesize(0.00005; 0.00007) & \footnotesize(0.00005; 0.00006) & \footnotesize(0.00004; 0.00005) & \footnotesize(0.00003; 0.00005)\\
\multirow{2}{*}{$H$} & 1.091 & 1.091 & 1.090 & 1.091 & 1.091\\
 & \footnotesize(1.0895; 1.0921) & \footnotesize(1.0887; 1.0939) & \footnotesize(1.0889; 1.0915) & \footnotesize(1.0892; 1.0929) & \footnotesize(1.0893; 1.0944)\\
\hline
\end{tabular}
\caption{Posterior summaries: median and 95\% credibility interval for the Log-Normal for years 1980, 1990, 2000, 2010 and 2019.}
\label{sec4:tab1}
\end{table}
One way to evaluate the behaviour of mortality for the United States population over the years and ages is by analysing the estimate of the HP parameters for the five years selected, as shown in Table  \ref{sec4:tab1}. \pkg{BayesMortalityPlus} provides tools to check the convergence of the generated Markov chains obtained in the estimation procedure (for more details, see \code{plot_chain}). To investigate the mortality improvement over five years, we consider assessing the significance of the parameters through the credible intervals criterion. We consider that there is a significant difference when the credible intervals are disjoint. Notice that parameter $A$ decreases significantly from 1980 to 1990 and from 1990 to 2000,  and then there is no significant difference, but the point estimate continues to decline. Parameters $B$ and $C$ show similar interpretations as parameter $A$,  both increasing over time until 2010, but in 2019 their estimates decrease to values close to the ones in 2000. It indicates that the changes are not statistically significant in a shorter temporal window.  In summary, the level of mortality in the first years of life decreased significantly over the years, except for the last year of the analysis.

For the second term of the Heligman-Pollard law, see that parameter $D$ decreases significantly over the years up to 2000, analogous to the estimates of parameter $A$. In 2019, there is a significant increase in its estimate. That means that the level of mortality in the accident hump decreased until 2000, persisted in this level in 2010, and increased in 2019. Parameter $E$ indicates that 1980 and 2000 were the years in which the mortality in the accident hump was most severe. The estimate of parameter $F$ is around 21 to 24 years until 2010 and in 2019 its estimate becomes almost 29 years, indicating a large shift of the accident hump to older ages.
For the last term of the HP function, parameter $G$ shows a decreasing behaviour over the years, which means that the level of mortality in adulthood  is decreasing over the years. This reduction was significant in the period from 2000 to 2010. On the other hand, parameter $H$ remains almost constant in all fits. 

\begin{figure}[t!]
    \centering
    \includegraphics[width = 0.8\textwidth]{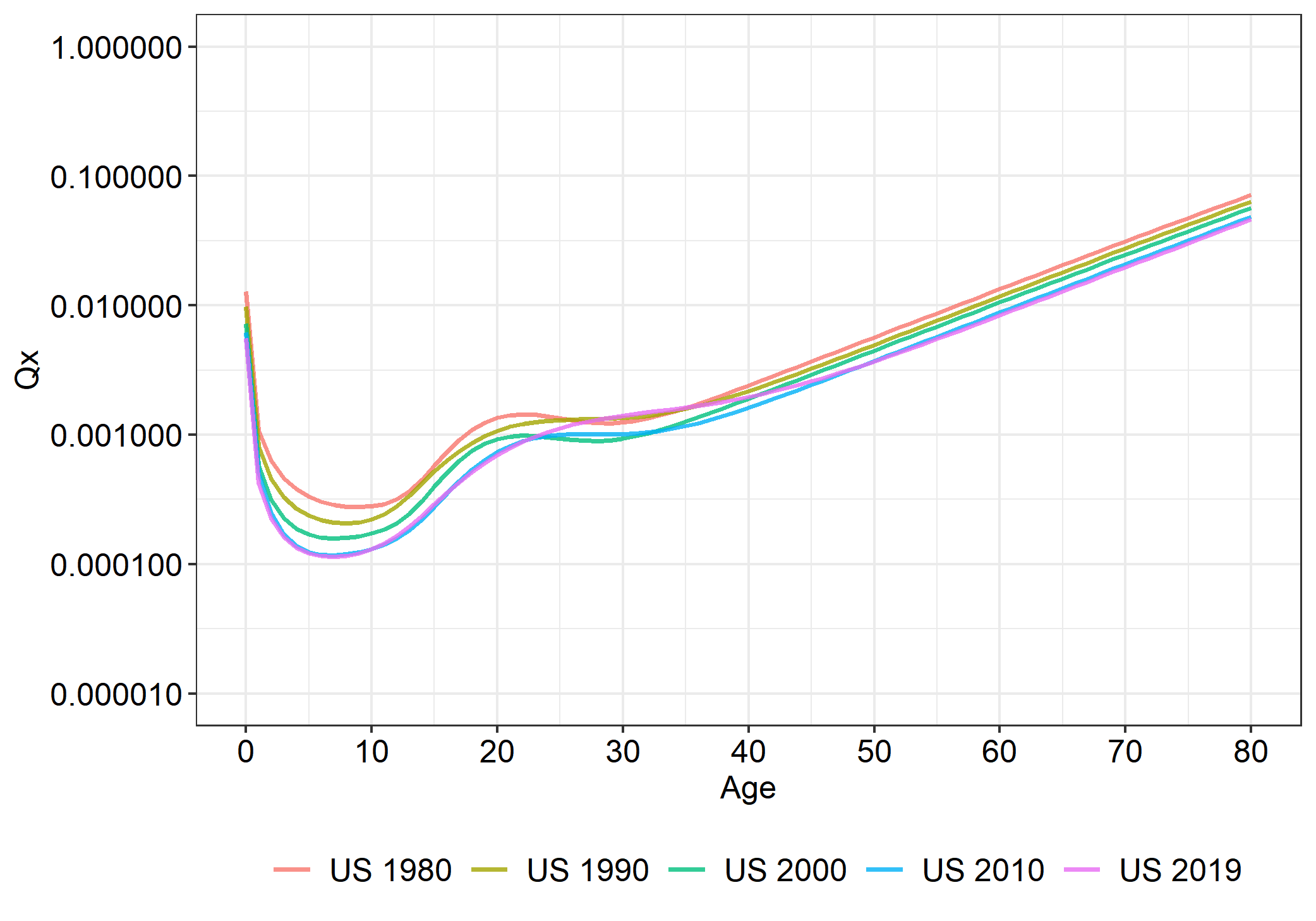}
    \caption{Posterior summaries via HP: median mortality curve in log-scale. The United States, total population, ages 0-80, and years 1980, 1990, 2000, 2010, 2019.}
    \label{fig:fitted_xables}
\end{figure}

To facilitate comparison among the five fitted models, we access the function \code{plot} available on \proglang{R} to visualize the behaviour of the mortality curves.  Figure \ref{fig:fitted_xables} shows the fitted mortality curves for the United States population. 
\begin{CodeChunk}
\begin{CodeInput}
R> fits <- list(fit_1980,fit_1990,fit_2000,fit_2010,fit_2019)
R> labels <- c("US 1980","US 1990","US 2000","US 2010", "US 2019")
R> plot(fits, labels = labels, plotData = F, plotIC = F)
\end{CodeInput}
\end{CodeChunk}
As seen in Figure \ref{fig:fitted_xables}, there is a similar behaviour of the adjusted tables in the first four years of the analysis, except for some level changes that occur between 1980 and 1990 in ages $x=25$ and $x=34$ and also between 2000 and 2010, in ages $x=24$ and $x=32$. This indicates consistency in the mortality pattern in the US population until 2010. On the other hand, this pattern is missed in 2019, where it can be observed that the accident hump is longer than in previous years, indicating that the causes of death that make up the accident hump are lasting longer than they used to.

\vspace{0.5cm}
\noindent {\it Predictive Credible Interval for the probability of death} 

\begin{figure}[t!]
    \centering
    \includegraphics[width = 0.8\textwidth]{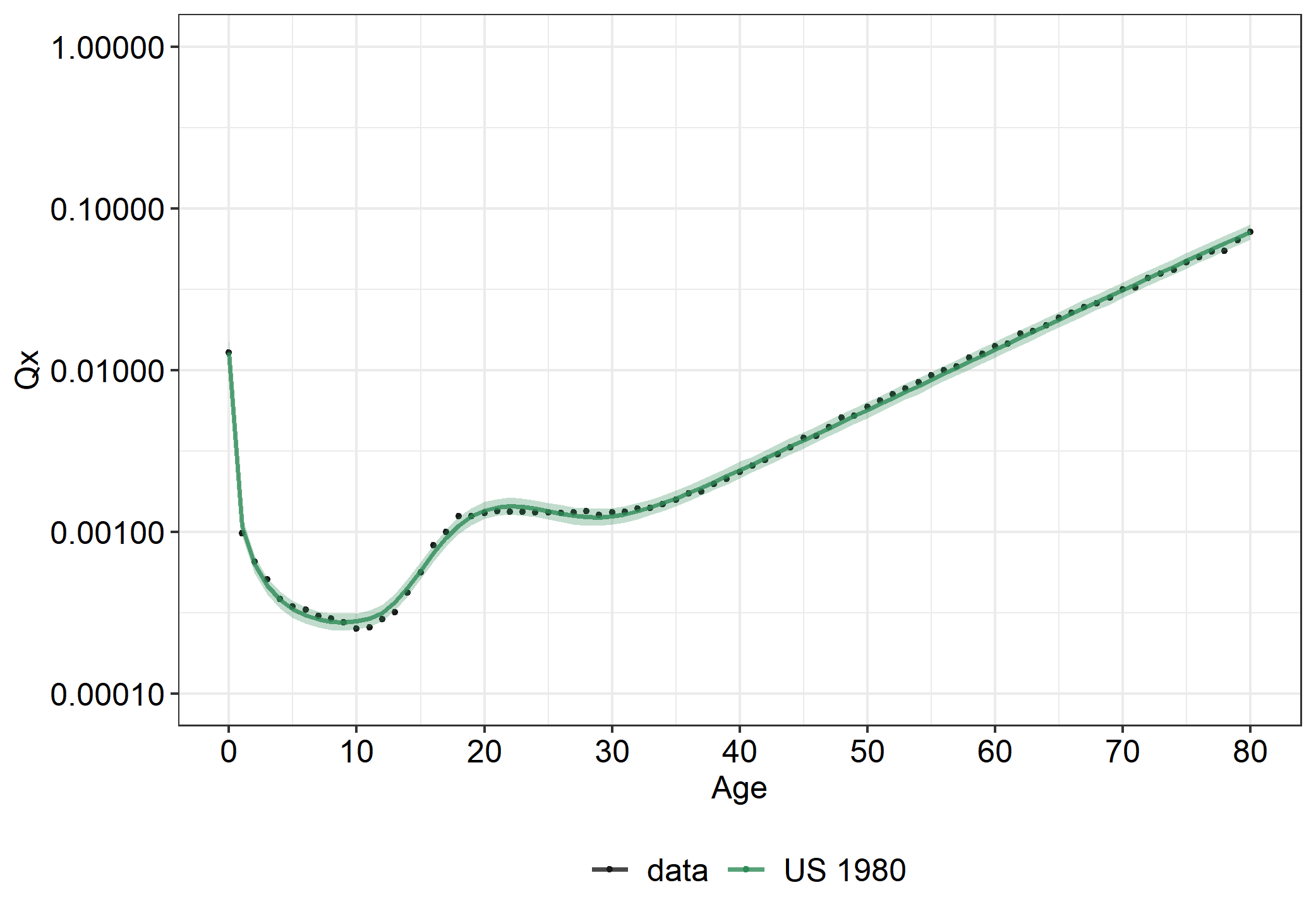}
    \caption{Posterior summaries via HP: median mortality curve and  95\% predictive credible interval in log-scale. The United States, total population, ages 0-80, and year 1980. The black dots represent the raw mortality rates. }
    \label{fig:plot_CI}
\end{figure}

Function \code{qx_ci} computes the predictive credible interval for $q_x$ from \code{hp} and \code{dlm} objects via the composition sampling technique \citep[see Chapter 5 of][]{gelfand04}. The following code provides credible intervals based on the HP fit for the year 1980:
\begin{CodeChunk}
\begin{CodeInput}
R> head(qx_ci(fit_1980, age= 1:81, Ex= NULL, prob=0.95))
  age           qi           qs
1   1 0.0009389319 0.0012430310
2   2 0.0005547195 0.0007122412
3   3 0.0004101589 0.0005214704
4   4 0.0003343556 0.0004244586
5   5 0.0002953454 0.0003724304
6   6 0.0002702347 0.0003420523
\end{CodeInput}
\end{CodeChunk}
Arguments \code{fit} and \code{age} are the same as defined in function \code{fitted}. Parameter \code{Ex} is a vector of the exposures that is used when the Binomial and Poisson models are fitted since both depend on these quantities. By default, \code{age} and \code{Ex} are set to be the ones passed in the fitting function. If any age outside of those used in the fitted curve is specified, the exposure for that age must also be determined by the user. This is not applied when the Log-Normal model is fitted. Additionally, the user can specify the probability of the predictive credible interval through the argument \code{prob}. 
Figure \ref{fig:plot_CI} presents the fitted mortality curve with the 95\% predictive credible interval. Figure \ref{fig:plot_CI} was obtained with the code: 
\begin{CodeChunk}
\begin{CodeInput}
R> plot(fit_1980, labels = "US 1980", plotIC = T, plotData = T)
\end{CodeInput}
\end{CodeChunk}

\noindent {\it Life Expectancy} 

\begin{figure}[H]
    \centering
    \includegraphics[width = 0.8\textwidth]{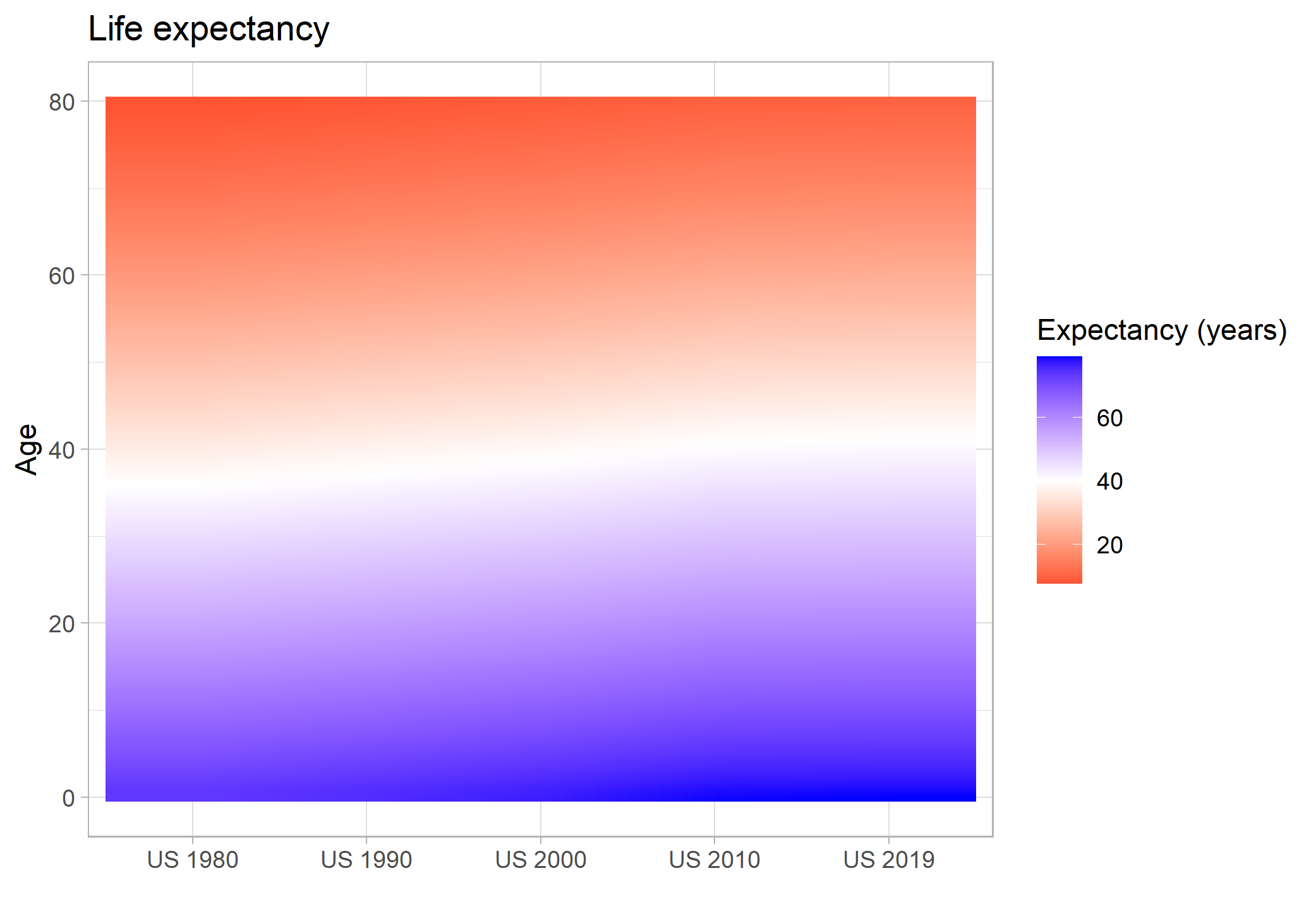}
    \caption{Posterior summaries via HP: life expectancy for the United States. Total population, 0-80 age for 1980, 1990, 2000, 2010, and 2019.}
    \label{fig:heatmap}
\end{figure}

The estimate of life expectancy is obtained via function \code{expectancy} through the curtate life expectancy as follows
\begin{equation}
e_x = \sum_{k=1}^\omega {}_{k}p_x,
\end{equation}
where $\omega$ is the maximum age available in the life table, and $_{k}p_{x}$ is the probability that someone aged $(x)$ will attain age $x+k$. Under the assumption of age-independent mortality, we can write $_{k}p_{x}$ as a cumulative product in terms of the survival probability $_{k}p_{x} = \prod_{i=0}^{k-1}p_{x+i} = p_x p_{x+1} p_{x+2} \ldots p_{x+k-1}$. Then,
\begin{equation}
    e_x = \sum_{k=1}^\omega \left(\prod_{i=0}^{k-1} p_{x+i}\right).
\end{equation}
The function is called in thee package as follows:
\begin{Code}
expectancy(fit, Ex = NULL, age = NULL, graph = TRUE, max_age = 110,
           prob = 0.95)   
\end{Code}
The output from function \code{expectancy} is an object of the class \code{hp} and \code{dlm}, and its arguments are: 
\begin{itemize}
\item \code{fit} represents the fitted curve by Heligman-Pollard or Dynamic Linear Model.
\item Arguments \code{Ex} and \code{prob} are exposure and probability, necessary to calculate the predictive intervals for the expectancy.
\item  By default, \code{Ex} is set to \code{NULL}  which indicates that the exposure available for the life expectancy is the same as used in the fitted curve. It is important to note that  argument \code{Ex} is used by the HP Binomial and HP Poisson models to associate the uncertainty with the ammount of available information.
\item Argument \code{max_age} (default=110) represents the maximum age to calculate the life expectancy. If necessary, the \code{expectancy()} function will extrapolate the fitted HP curve until it reaches the maximum age argument. In these cases, it is important to attend to \code{Ex} argument: if it is set to \code{NULL}, the function will repeat the last informed exposure to match the age interval.
\end{itemize}
The user can obtain the residual life expectancy for specific ages, with age 0 meaning life expectancy at birth. For illustration, consider setting the ages \code{0,20,40,60,80} and year 1980, with the following code:
\begin{CodeChunk}
\begin{CodeInput}
R> expectancy(fit_1980, age = c(0,20,40,60,80), graph = F)
   Age Expectancy Lower CI Upper CI
1    0      73.31    71.77    74.82
21  20      54.91    53.58    56.24
41  40      36.26    35.05    37.47
61  60      19.42    18.44    20.41
81  80       7.51     6.94     8.11
\end{CodeInput}
\end{CodeChunk}
Table \ref{tab:life_expectancy}  presents a summary of the posterior life expectancy for the five models and Figure \ref{fig:heatmap} illustrates the behaviour of the posterior distribution for the life expectancy for selected years. For all ages, we see an increase in life expectancy. Notice that the increase is not significant in close years but is remarkable when the decades are considered. In 2019,  the life expectancy at birth is superior by more than five years to that in 1980. In terms of a point estimates, the most considerable difference occurs between 2000 and 2010, while the smallest difference is between 2010 and 2019. 

In addition, \pkg{BayesMortalityPlus} package allows graphical visualization of the behaviour of life expectancy over the years. In Figure \ref{fig:heatmap}, we can see stronger blue tones at the bottom for the first ages throughout the years, which represents a larger life expectancy at birth. It is also noticeable that the ages with lightest tones are associated with a life expectancy close to 40 years, and have a slight increase over the years. Because in older ages the life expectancy is small and therefore becomes close for all years, there is no significant change at the top of the graphic. 
\begin{table}[t!]
\centering
\caption{Posterior summaries: life expectancy and 95\% interval credibility for the Log-Normal model for the US population at ages: 0, 10, 20, 30, 40, 50, 60, 70, 80.}
\label{tab:life_expectancy}
\begin{tabular}{lccccc}
\hline
Age & US 1980 & US 1990 & US 2000 & US 2010 & US 2019\\
\hline
\multirow{2}{*}{0} & 73.1 & 75.01 & 76.82 & 78.81 & 79.05\\
 & (71.76; 74.82) & (72.75; 77.24) & (75.23; 78.40) & (76.53; 81.04) & (76.89; 81.21)\\
\multirow{2}{*}{10} & 64.54 & 65.97 & 67.54 & 69.42 & 69.61\\
 & (63.17; 65.90) & (63.9; 68.03) & (66.05; 69.01) & (67.29; 71.54) & (67.57; 71.67)\\
\multirow{2}{*}{20} & 54.91 & 56.28 & 57.80 & 59.62 & 59.81\\
 & (53.57; 56.24) & (54.27; 58.31) & (56.35; 59.25) & (57.52; 61.71) & (57.8; 61.84)\\
\multirow{2}{*}{30} & 45.59 & 46.94 & 48.30 & 50.14 & 50.39\\
 & (44.32; 46.86) & (45.02; 48.87) & (46.91; 49.70) & (48.12; 52.15) & (48.47; 52.35)\\
\multirow{2}{*}{40} & 36.26 & 37.62 & 38.85 & 40.68 & 41.14\\
 & (35.04; 37.47) & (35.79; 39.47) & (37.51; 40.20) & (38.74; 42.62) & (39.31; 43)\\
\multirow{2}{*}{50} & 27.41 & 28.68 & 29.83 & 31.54 & 32.08\\
 & (26.28; 28.54) & (26.99; 30.41) & (28.58; 31.10) & (29.72; 33.39) & (30.37; 33.84)\\
\multirow{2}{*}{60} & 19.42 & 20.52 & 21.57 & 23.09 & 23.59\\
 & (18.43; 20.41) & (19.03; 22.07) & (20.46; 22.72) & (21.45; 24.77) & (22.05; 25.20)\\
\multirow{2}{*}{70} & 12.67 & 13.55 & 14.45 & 15.69 & 16.11\\
 & (11.87; 13.48) & (12.33; 14.85) & (13.52; 15.41) & (14.32; 17.12) & (14.8; 17.48)\\
\multirow{2}{*}{80} & 7.51 & 8.14 & 8.83 & 9.74 & 10.04\\
 & (6.93; 8.1) & (7.23; 9.12) & (8.13; 9.55) & (8.69; 10.85) & (9.03; 11.12)\\
\hline
\end{tabular}
\end{table}
\begin{CodeChunk}
\begin{CodeInput}
R> Heatmap(fits, x_lab = labels, age = 0:80)
\end{CodeInput}
\end{CodeChunk}

\noindent {\it Modelling adult ages}\\

\begin{figure}[ht!]
    \centering
    \includegraphics[width = 0.8\textwidth]{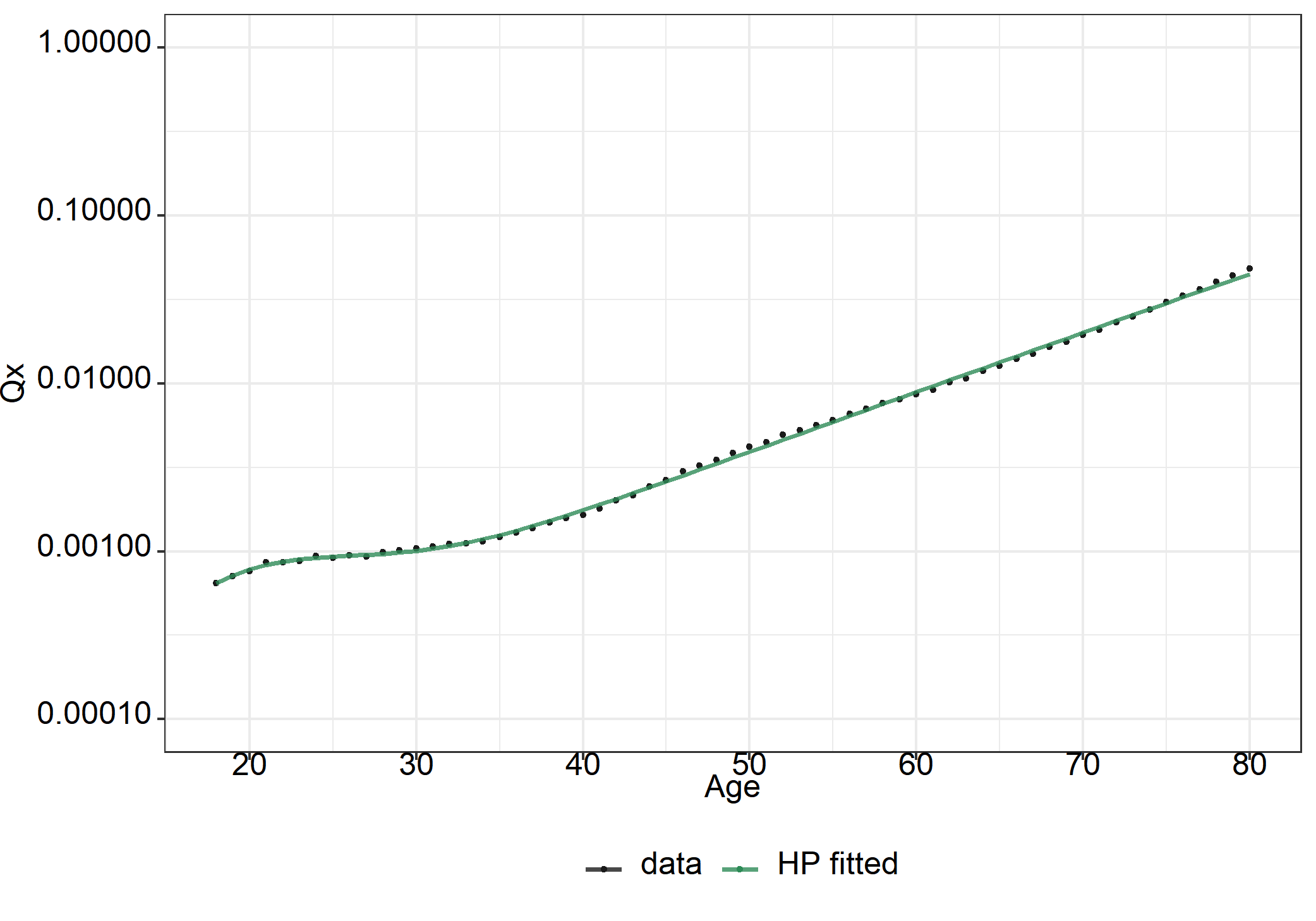}
    \caption{Posterior summaries via HP: median mortality curve in log-scale via the reduced model. The United States, total population, ages 18-80, the year 2010. Black dots represent the raw mortality rates.}
    \label{fig:reduced_fit}
\end{figure}

Assume a scenario in which interest lies in modelling only the mortality rate for adults. Some issues could result in this scenario such as poor quality of the data on the mortality of infants and young ages or their non-existence. For example, consider an insurance life product for employees of a company. In this case, we would not access data for children and young people due to the fact that they are not legally allowed to work. In this case, the younger ages are not modelled as the first term of the HP curve is not available for analysis. The argument \code{reduced\_model} is available in function \code{hp} to deal with such situations. For illustration, consider the data for the year 2010 from age 18 (see Figure \ref{fig:reduced_fit}). 
\begin{CodeChunk}
\begin{CodeInput}
R> hp.fit2 <- hp(x = 18:80, Ex = ex_2010[19:81], Dx = dx_2010[19:81],
                model = "lognormal", reduced_model = TRUE)
R> plot(hp.fit2, plotIC = F, labels = "HP fitted")
\end{CodeInput}
\end{CodeChunk}
This approach allows fast computations and convergence in the MCMC algorithm for two reasons. Firstly, the dimension of the parameters is reduced, the acceptance rate raises and consequently, the algorithm converges fast. Furthermore, identifiability issues in the parameters $A$, $B,$ and $C$ could occur if they are taken into account in the inference procedure. These parameters are not estimated when considering the \code{reduce\_model} argument.\\

\newpage
\noindent {\it Mortality measurement at advanced ages and extrapolation}\\

The life tables are composed of the mortality probability $q_x$, associated with each age $x$. The estimates of mortality at advanced ages are difficult to compute due to the fact that there is a small number of survivors in this age group. In this context, to achieve a robust fit at advanced ages,  following  \cite{hustead2005ending} we consider four methodologies to accommodate the mortality pattern at the end of the mortality graduation.
The synopsis of the function \code{hp\_close} is given by:
\begin{Code}
hp_close(fit, method = c("hp", "plateau", "linear", "gompertz"),
   x0 = max(fit$data$x), max_age = 120, k = 7,
   weights = seq(from = 0, to = 1, length.out = 2*k+1),
   new_Ex = NULL, new_Dx = NULL)
\end{Code}
This function receives an object of the class \code{"HP"} adjusted by the function \code{hp} and fits a closing method to expand the data of the life table to a maximum age argument \code{max_age} (default=120) inputted by the user.  The user can adopt alternative approaches for closing tables using the argument \code{method}. The package provides  four closing methods: \code{method="hp"}, \code{method="plateau"}, \code{method="linear"} and \code{method="gompertz"}. Notice that  \code{method= "linear"} can only be used with HP objects following \code{model="lognormal"} option of the HP model. Also,
\begin{itemize}
    \item \code{x0}, \code{k} and \code{weights} arguments control the mixture of the fitted HP model to the closing model, representing the starting age for the closing method, the size of the age interval to be mixed and the weights to be applied to the mixture, respectively.
    \item \code{new_Ex} and \code{new_Dx} arguments represent the data that was not fitted by the original HP curve, the exposure, and the death count after the \code{x0} argument. These arguments must be the same length and are required for the Binomial and Poisson mortality models, and also for \code{method="linear"} and \code{method="gompertz"}.
\end{itemize}

For illustration of the closing methods available in \pkg{BayesMortalityPlus} package, we consider the object \code{fit_2019} and age 80-100 (extrapolation).\\

\noindent {\it 1. The HP method}\\ 
The argument \code{method="hp"} extrapolates the fitted \cite{heligman1980age} curve to the \code{max_age} argument informed by the user. We expect the mean of the posterior distribution to be similar to the values obtained by the \code{fitted} function, as well as the predictive intervals obtained by the \code{qx_ci} function. Since it is just an extrapolation of the HP curve, the mixture is not applied.
\begin{CodeChunk}
\begin{CodeInput}
R> hp.close1 <- hp_close(fit_2019, method="hp", max_age = 100)
\end{CodeInput}
\end{CodeChunk}
%
%


\noindent {\it 2.  The Plateau method}\\
The \code{method= "plateau"} considers that the death probability $q_x$ of the last age fitted by the HP model is kept constant until it reaches the maximum age. No mixture is applied. More detailed discussion about this mortality pattern at the oldest ages can be seen in \cite{lai2012human}.
\begin{CodeChunk}
\begin{CodeInput}
R> hp.close2 <- hp_close(fit_2019, method="plateau", max_age = 100)
\end{CodeInput}
\end{CodeChunk}
%

\noindent {\it 3. The Linear method}\\
The \code{method="linear"} is only available for the Log-Normal model. This method fits a linear regression starting at age $x_0 - k$ until the last age with available data and is specified as:
\begin{equation}
    \begin{aligned}
        log(q_x) = \beta_0 + \beta_1 x + \varepsilon_x \\ 
        \varepsilon_x \sim N(0,\sigma_0^2)
    \end{aligned}
\end{equation}
After fitting the linear regression, predictive samples are generated for the death probabilities $q_x$ starting at age $x_0 - k$, dividing the graduation into three parts: the first one is the fitted curve given by the HP model, followed by the mixture interval given by the \code{k} argument, ending with the fitted linear regression.
\begin{CodeChunk}
\begin{CodeInput}
R> new_Ex <- dplyr::filter(USA, Year == 2019)$Ex.Total[81:101]
R> new_Dx <- dplyr::filter(USA, Year == 2019)$Dx.Total[81:101]
R> hp.close3 <- hp_close(fit_2019, method="linear", max_age = 100,
                          new_Ex = new_Ex, new_Dx = new_Dx)
\end{CodeInput}
\end{CodeChunk}
\noindent {\it 4. The Gompertz method}\\
Details about the  \code{method="gompertz"} are available in \cite{Forster2018} and \cite{gavrilov2011mortality}. \cite{Forster2018} consider the Gompertz curve to close the English life tables between 2010-2012 and conclude that the better the quality of mortality data at advanced ages, the more the behaviour of the mortality curve approaches the Gompertz function. 

\begin{figure}[!ht]
    \centering
    \includegraphics[width = 0.8\textwidth]{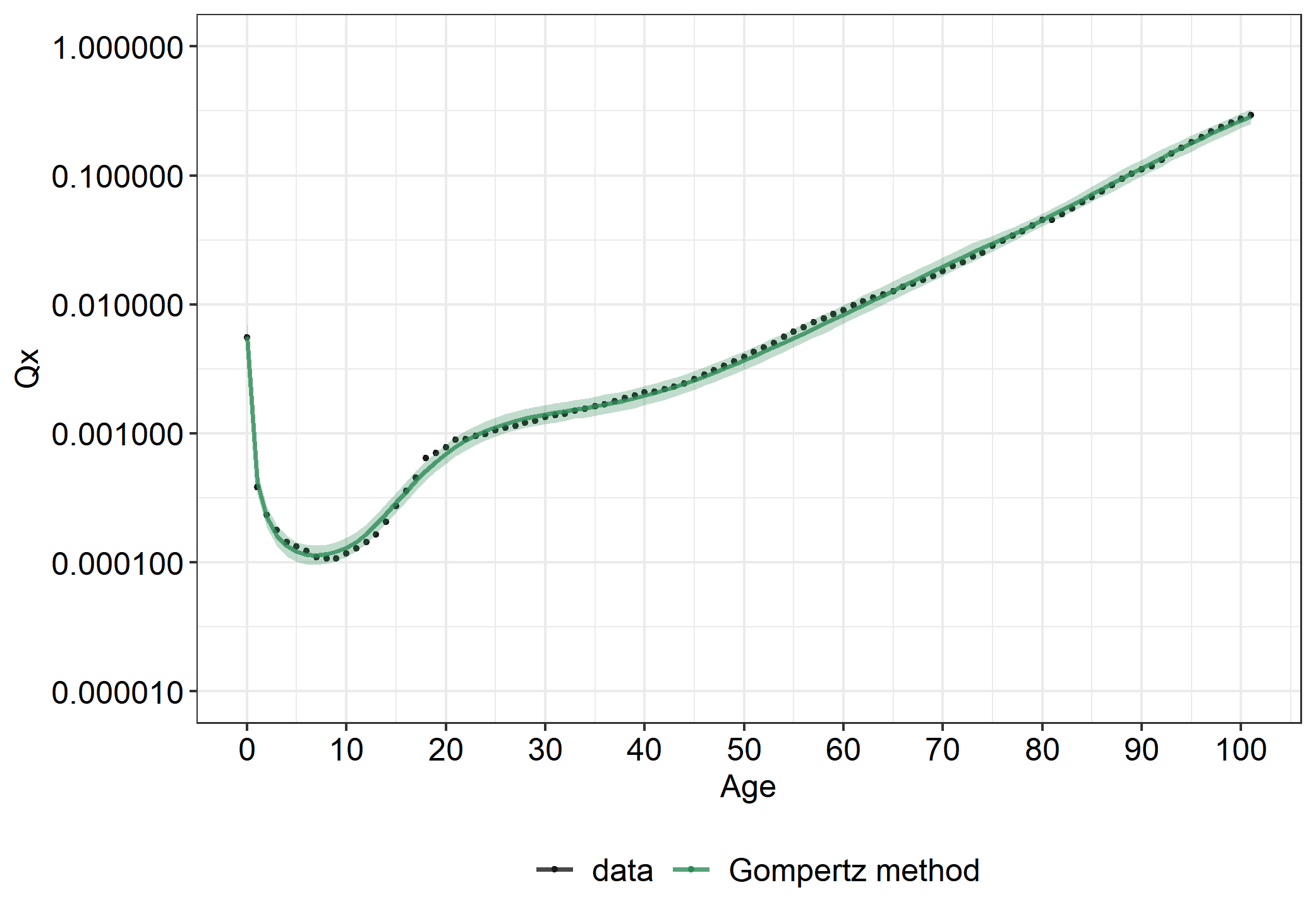}
    \caption{Posterior summaries via HP: mortality graduation with Gompertz closing method in log-scale. The United States, total population, ages 0-100 and year 2019.}
    \label{fig:example_gompertz}
\end{figure}

This method fits the Gompertz curve developed by \cite{gompertz1825xxiv} through the Sampling Importance Resampling method (SIR) as 
\begin{equation}
  \begin{aligned}
    \mu(x) = Ae^{Bx}  \\
    q_x = 1 - e^{-\mu(x)},
    \end{aligned}
\end{equation}
%
%
\noindent where $\mu(x)$ represents the mortality force that grows exponentially with the increase of age $x$, depending on parameters $A$ and $B$. Parameter $A \in (0, 1)$ reflects the general level of mortality, while $B \in (0, \infty)$ controls the rate at which the force of mortality increases with age. Notice that if we assume $C = e^{B}$, then $\mu(x) = A C^x$ that is equivalent to the third term of the eight-parameter HP curve previously seen in Section \ref{sec2}.
The results of the Gompertz method are plotted in Figure \ref{fig:example_gompertz} with  95\% predictive credible interval, for the year 2019. Analogously, we can plot the mortality curve and expectancy life for the objects \code{fit\_hp}, \code{fit\_plat}, and \code{fit\_lin}, respectively.
\begin{CodeChunk}
\begin{CodeInput}
R> hp.close4 <- hp_close(fit_2019, method="gompertz", max_age = 100,
                     new_Ex = new_Ex, new_Dx = new_Dx)
R> plot(hp.close4, labels = "Gompertz method")
\end{CodeInput}
\end{CodeChunk}

\begin{figure}[!ht]
    \centering
    \includegraphics[width = 0.8\textwidth]{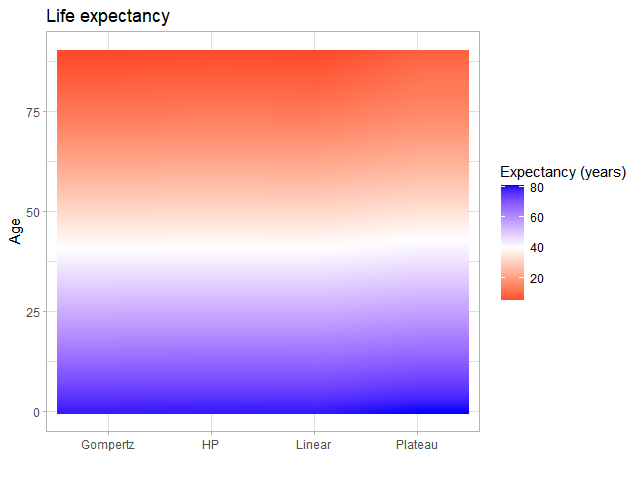}
    \caption{Posterior summaries via HP: life expectancy of each closing method for the United States. Total population, 0-100 age and year 2019.}
    \label{fig:heatmap_close}
\end{figure}

After choosing the closing method, an \code{`ClosedHP'} object will be generated to save the new life table. This allows the new complete graduation to be used as an argument to other functions in the package.  For example, in Figure \ref{fig:heatmap_close} the life expectancy is shown and compared with different closing methods via object \code{fits.new}. 
\begin{CodeChunk}
\begin{CodeInput}
R> fits.new <- list(hp.close1, hp.close2, hp.close3, hp.close4)
R> Heatmap(fits.new, x_lab = c("HP","Plateau","Linear","Gompertz"))
\end{CodeInput}
\end{CodeChunk}

\subsection[Mortality Graduation via Dynamical Linear Smoothers]{Mortality Graduation via Dynamical Linear Smoothers}\label{sec4.2}
\begin{figure}[!ht]
    \centering
    \includegraphics[width = 0.8\textwidth]{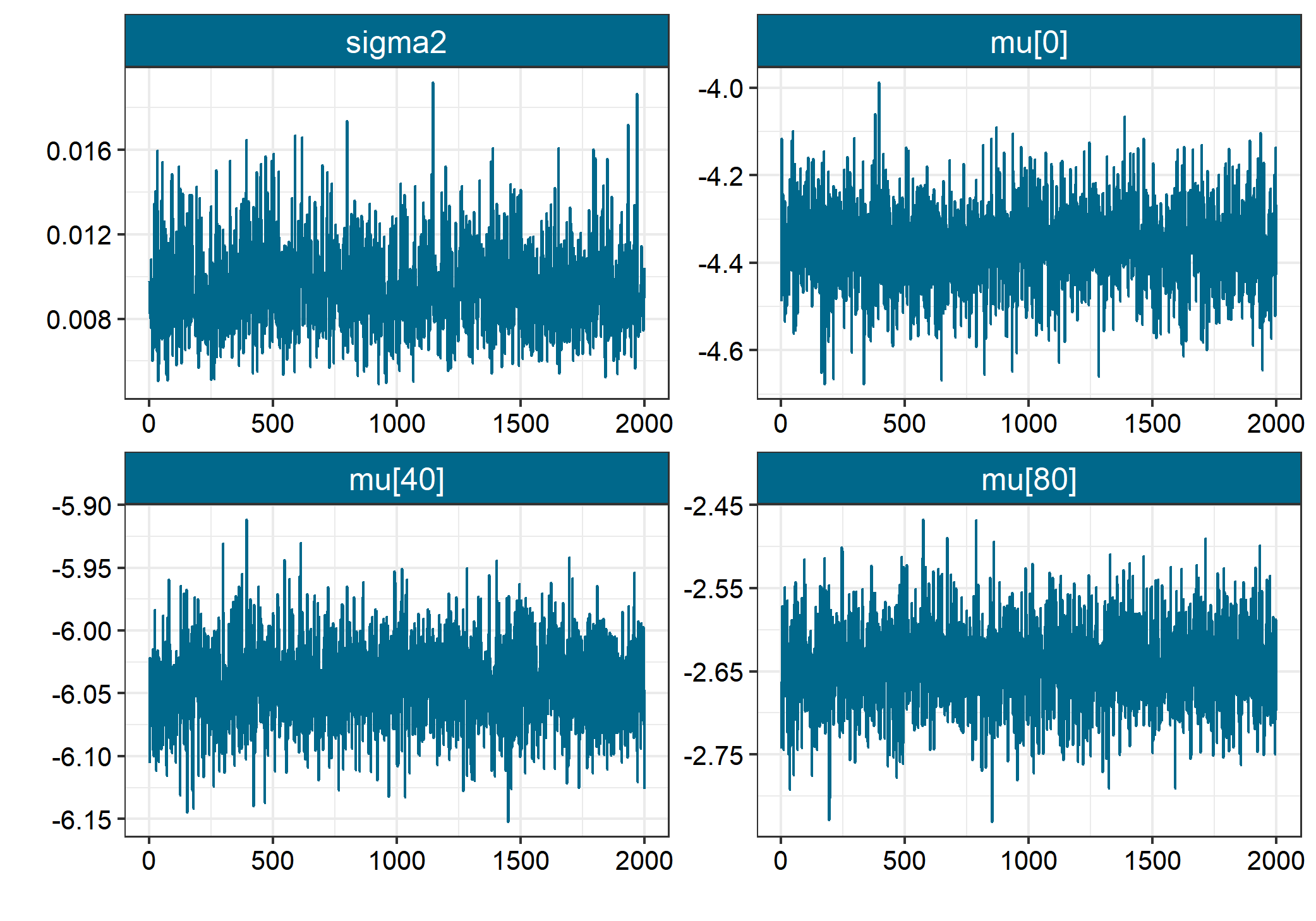}
     \caption{Posterior summaries via DLM: traces of the chains for the mean $\mu_x$ ($x=0, 40, 80)$ and variance $V$ of the process. The United States, total population, and year 1980. }
    \label{fig:chains_dlm}
\end{figure}

\begin{figure}[!ht]
    \centering
    \includegraphics[width = 0.8\textwidth]{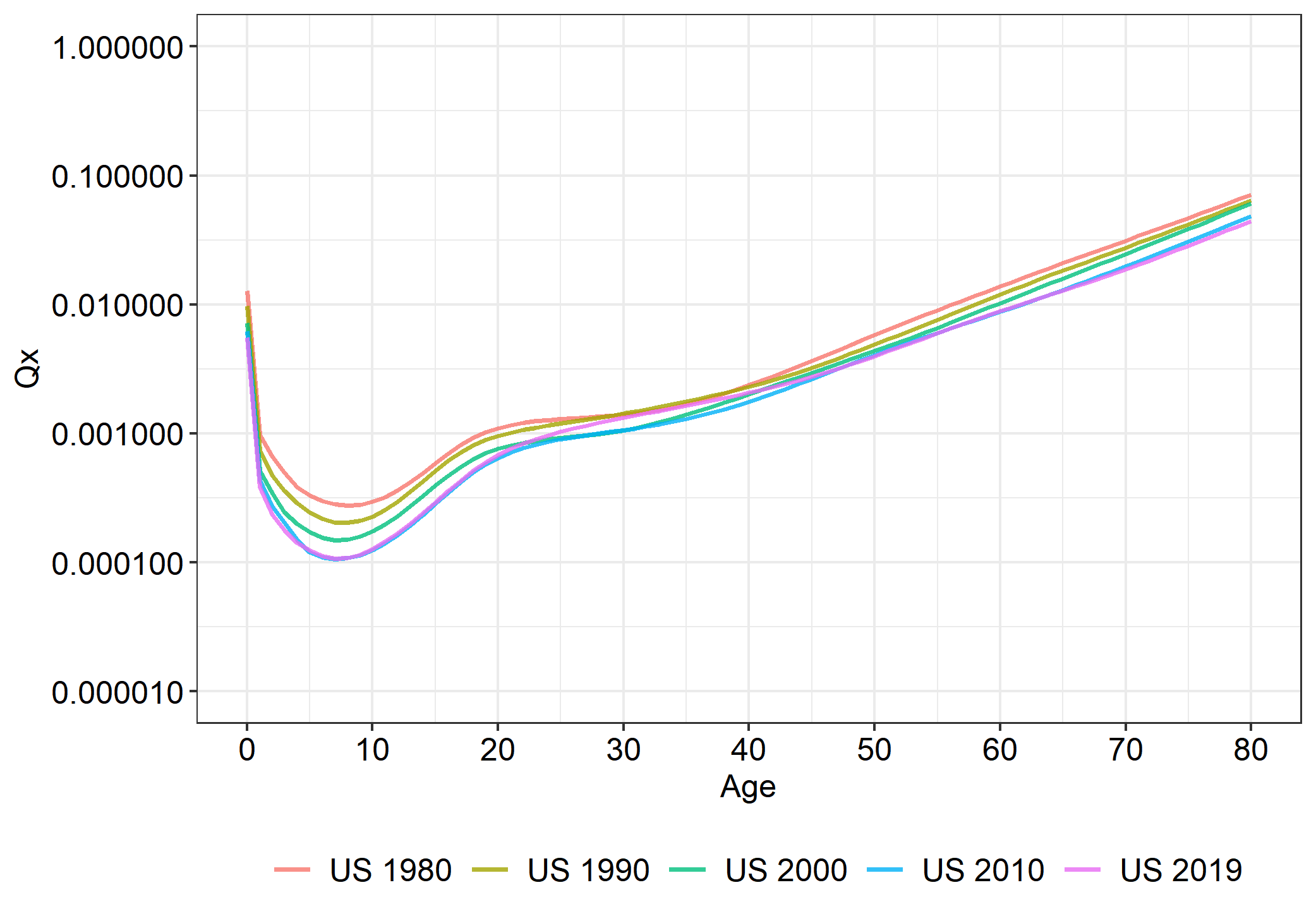}
     \caption{Posterior summaries via DLM: median mortality curve in log-scale. The United States, total population, ages 0-80, and years 1980, 1990, 2000, 2010, 2019.}\label{fig:plot_dlm}
\end{figure}

The function \code{dlm} returns an object of class \code{"DLM"}, which is a Dynamic linear model with input data settled by the user. We consider the same dataset for which we have applied the Heligman-Pollard model,  with the data being transformed into log mortality to reproduce the results of the graduation mortality curves. To fit a dynamical linear model under the log mortality for five different years, consider the following code:
\begin{CodeChunk}
\begin{CodeInput}
R> y_1980 <- log(dx_1980/ex_1980)
R> y_1990 <- log(dx_1990/ex_1990)
R> y_2000 <- log(dx_2000/ex_2000)
R> y_2010 <- log(dx_2010/ex_2010)
R> y_2019 <- log(dx_2019/ex_2019)

R> dlm_1980 <- dlm(y_1980, delta=0.85)
Simulating [===================================] 100% in 38s
R> dlm_1990 <- dlm(y_1990, delta=0.85)
Simulating [===================================] 100% in 37s
R> dlm_2000 <- dlm(y_2000, delta=0.85)
Simulating [===================================] 100% in 43s
R> dlm_2010 <- dlm(y_2010, delta=0.85)
Simulating [===================================] 100% in 42s
R> dlm_2019 <- dlm(y_2019, delta=0.85)
Simulating [===================================] 100% in 41s    
\end{CodeInput}
\end{CodeChunk}
Posterior summaries for the five fitted models are allowed via the \code{summary} function available in \code{R}. For illustration, we exhibit posterior summaries for the year 1980 and some ages, as follows:
\begin{CodeChunk}
\begin{CodeInput}
R> head(summary(dlm_1980))

           mean      sd     2.5%    50.0%    97.5%
sigma2  0.00932 0.00203  0.00594  0.00910  0.01380
mu[0]  -4.35763 0.09607 -4.55490 -4.35698 -4.15945
mu[1]  -6.92630 0.09633 -7.11315 -6.92766 -6.74431
mu[2]  -7.32963 0.09376 -7.52254 -7.32660 -7.14864
mu[3]  -7.60248 0.09030 -7.78687 -7.60194 -7.43082
mu[4]  -7.86710 0.07516 -8.01933 -7.86680 -7.71978
\end{CodeInput}
\end{CodeChunk}
\noindent Note that \code{mu[]} represents the posterior mean of the log mortality rate for each age in the study. The user can also call the \code{plot_chain} function to visualise the traces of the generated chains for the estimated parameters. 
The traces of the chains are plotted discarding the burn-in period necessary to achieve convergence and taking into account the thinning for elimination of serial autocorrelation. Details about the default values or specification of these quantities, as well as of the total number of iterations considered in the MCMC algorithm, can be examined by referring respectively to arguments \code{bn},  \code{thin} and  \code{M}, in the fitting function \code{dlm} (and \code{hp}). 
As an illustration, see in Figure \ref{fig:chains_dlm} the traces for the posterior chains (based on data from 1980) for the mean and variance of the log mortality for ages 0, 40 and 80, respectively. Analogously, the user can resort to the same function to plot the posterior chains under the HP model. 
\begin{CodeChunk}
\begin{CodeInput}
R> plot_chain(dlm_1980, param=c("sigma2","mu[0]","mu[40]","mu[80]"))
\end{CodeInput}
\end{CodeChunk}
Figure \ref{fig:plot_dlm} shows the mortality curve fit via DLM graduation for the five years considered in the analysis, plotted through the function \code{plot} available on \proglang{R}.

\begin{figure}[!ht]
    \centering
    \includegraphics[width = 0.8\textwidth]{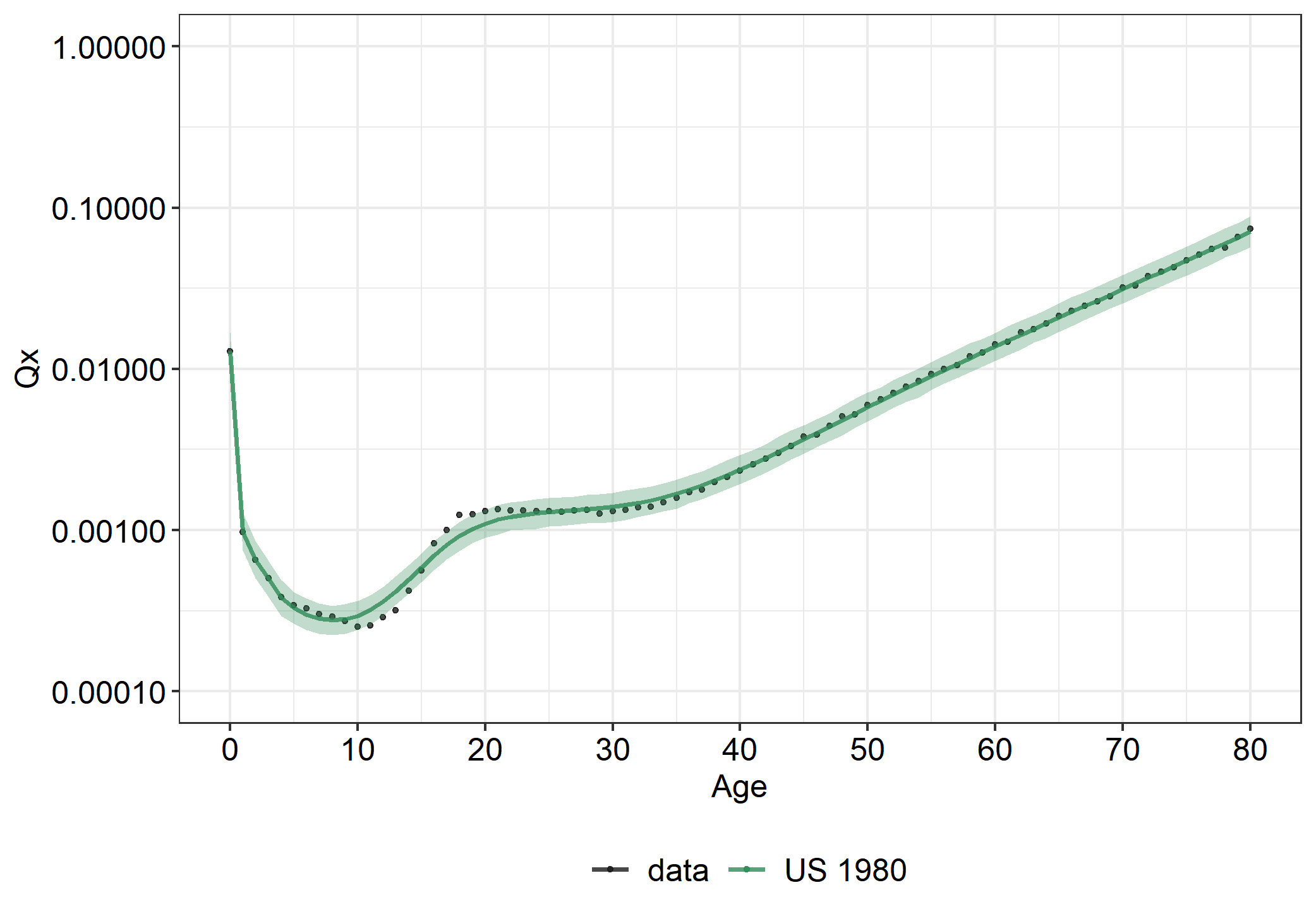}
      \caption{Posterior summaries via DLM: median mortality curve and predictive credible interval 95\% in log-scale. The United States, total population, ages 0-80, and year 1980. The black dots represent the raw mortality
rates.}
    \label{fig:ci_dlm}
\end{figure}

\begin{CodeChunk}
\begin{CodeInput}
R> fits <- list(dlm_1980,dlm_1990,dlm_2000,dlm_2010,dlm_2019)
R> labels <- c("US 1980","US 1990","US 2000","US 2010","US 2019")
R> plot(fits, labels = labels, plotData = F, plotIC = F)
\end{CodeInput}
\end{CodeChunk}
Predictive credible intervals for the mortality curves can be addressed using function \code{qx_ci}, already mentioned in Section \ref{sec4.1}. Figure \ref{fig:ci_dlm} presents the fitted mortality curve via DLM with the 95\% credible interval for the year 1980. 
\begin{CodeChunk}
\begin{CodeInput}
 R> plot(dlm_1980, labels = "US 1980", plotIC = T, plotData = T)
\end{CodeInput}
\end{CodeChunk}
As shown in Section \ref{sec4.1}, other posterior measures of the behaviour of the population can be computed. For example, life expectations for the ages informed by the user, as seen below: 

\newpage

\begin{CodeChunk}
\begin{CodeInput}
R> expectancy(dlm_1980, age = c(0,20,40,60,80), graph = F)

   Age Expectancy Lower CI Upper CI
1    0      73.19    70.26    76.34
21  20      54.77    52.22    57.62
41  40      36.10    33.76    38.78
61  60      19.32    17.36    21.66
81  80       7.41     5.96     9.45

R> Heatmap(fits, x_lab = labels, age = 0:80)
\end{CodeInput}
\end{CodeChunk}

\begin{figure}[!ht]
    \centering
    \includegraphics[width = 0.8\textwidth]{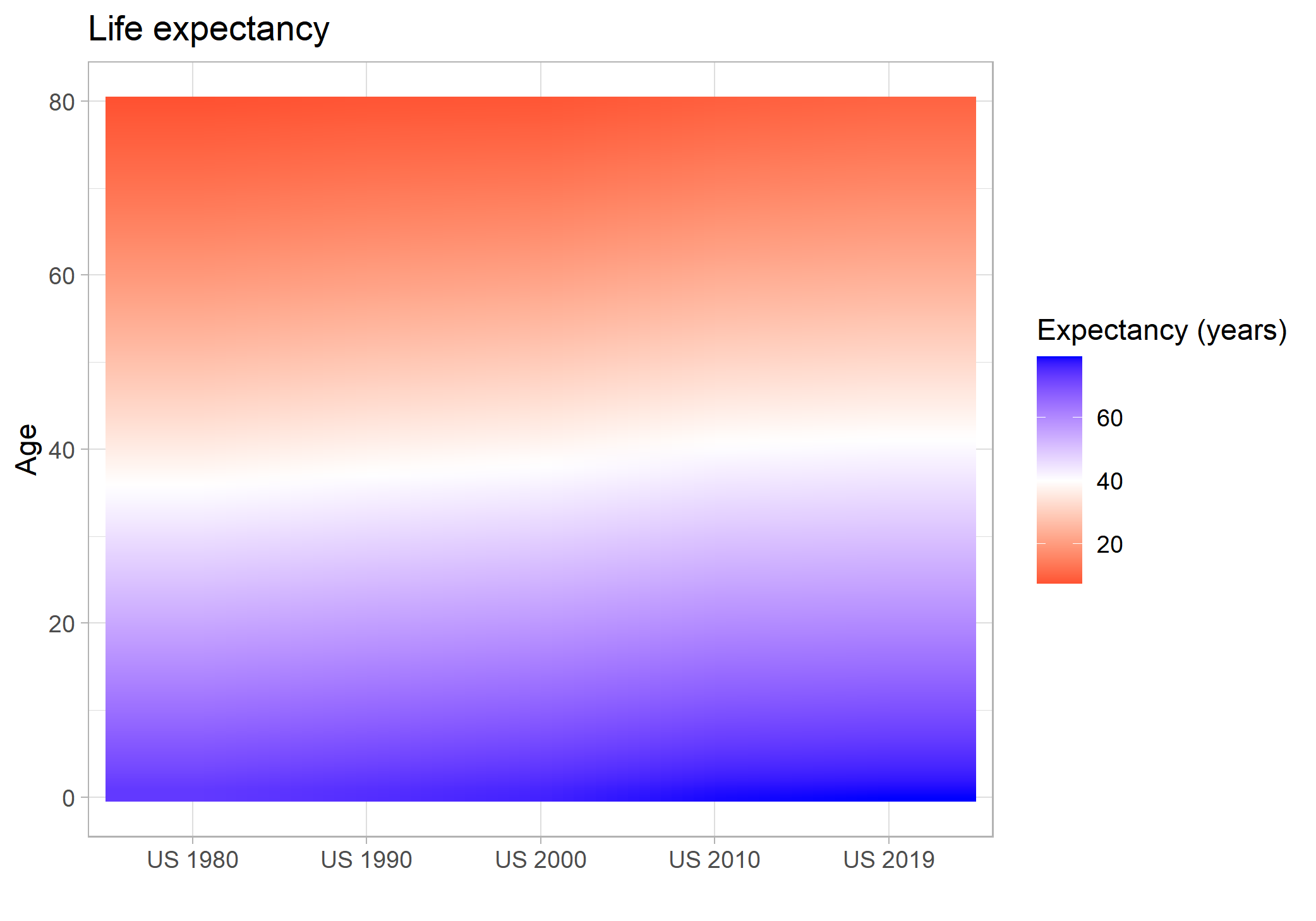}
    \caption{Posterior summaries via DLM: the life expectancy for the United States. Total population, 0-80 age for 1980, 1990, 2000, 2010, and 2019.}
    \label{fig:heatmap_dlm}
\end{figure}

\begin{figure}[!ht]
    \centering
    \includegraphics[width = 0.8\textwidth]{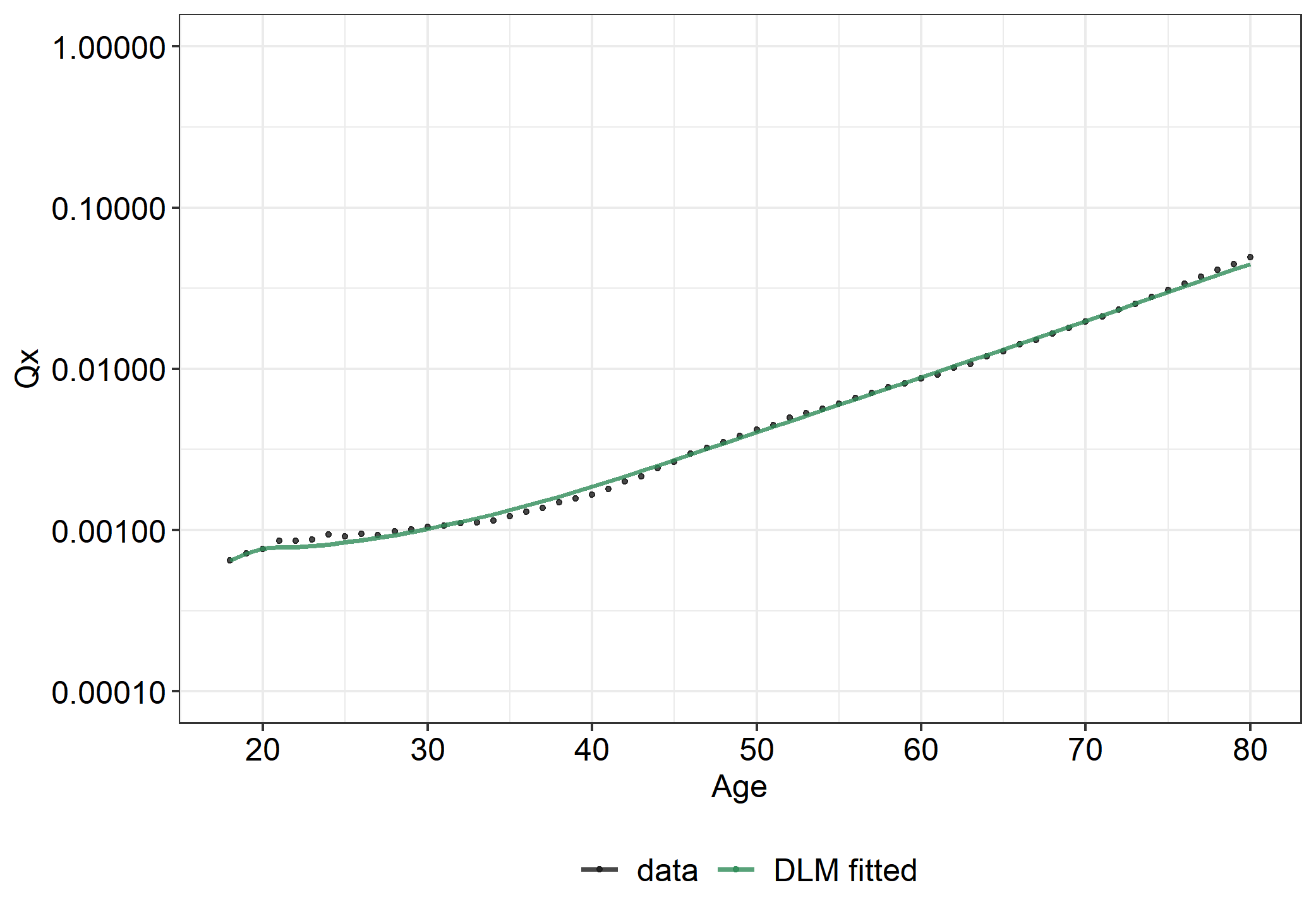}
      \caption{Posterior summaries via DLM: median mortality curve in log-scale to adult ages. The United States, total population, ages 18-80, the year 2010. Black dots represent the raw mortality rates.}
    \label{fig:red_dlm}
\end{figure}

Opposed to the Heligman-Pollard model, the DLM approach does not assume a parametric structure in terms of mortality laws, resulting in more flexibility in the table graduation. We can model the adult ages as seen previously in Section \ref{sec4.1}, through the \code{dlm} function, taking into account a range of the adult ages of interest. For illustration purposes, an age range from 18 to 80 is considered. The results are plotted in Figure \ref{fig:red_dlm}. 
\begin{CodeChunk}
\begin{CodeInput}
R> dlm.fit2 <- dlm(y_2010[19:81], delta=0.95, ages = 18:80)
Simulating [===================================] 100% in 34s
R> plot(dlm.fit2, plotIC = F, labels = "DLM fitted")
\end{CodeInput}
\end{CodeChunk}

{\it Mortality measurement at advanced ages and extrapolation}

\begin{figure}[!ht]
    \centering
    \includegraphics[width = 0.8\textwidth]{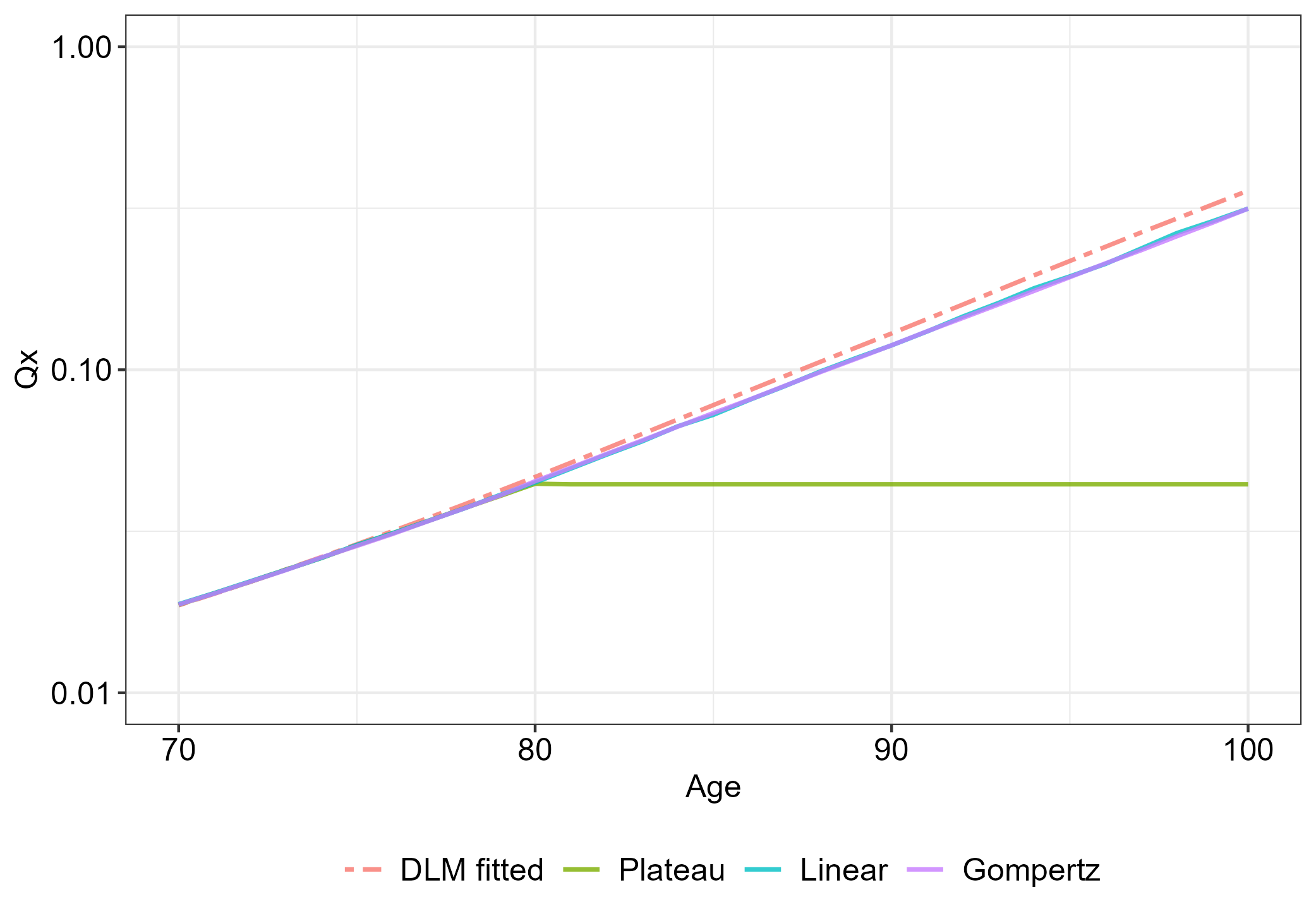}
      \caption{Posterior summaries via DLM: mortality graduation with different closing methods in log-scale. The United States, total population, ages 70-100, the year 2019.}
    \label{fig:closed_dlm}
\end{figure}

\begin{figure}[!ht]
    \centering
    \includegraphics[width = 0.8\textwidth]{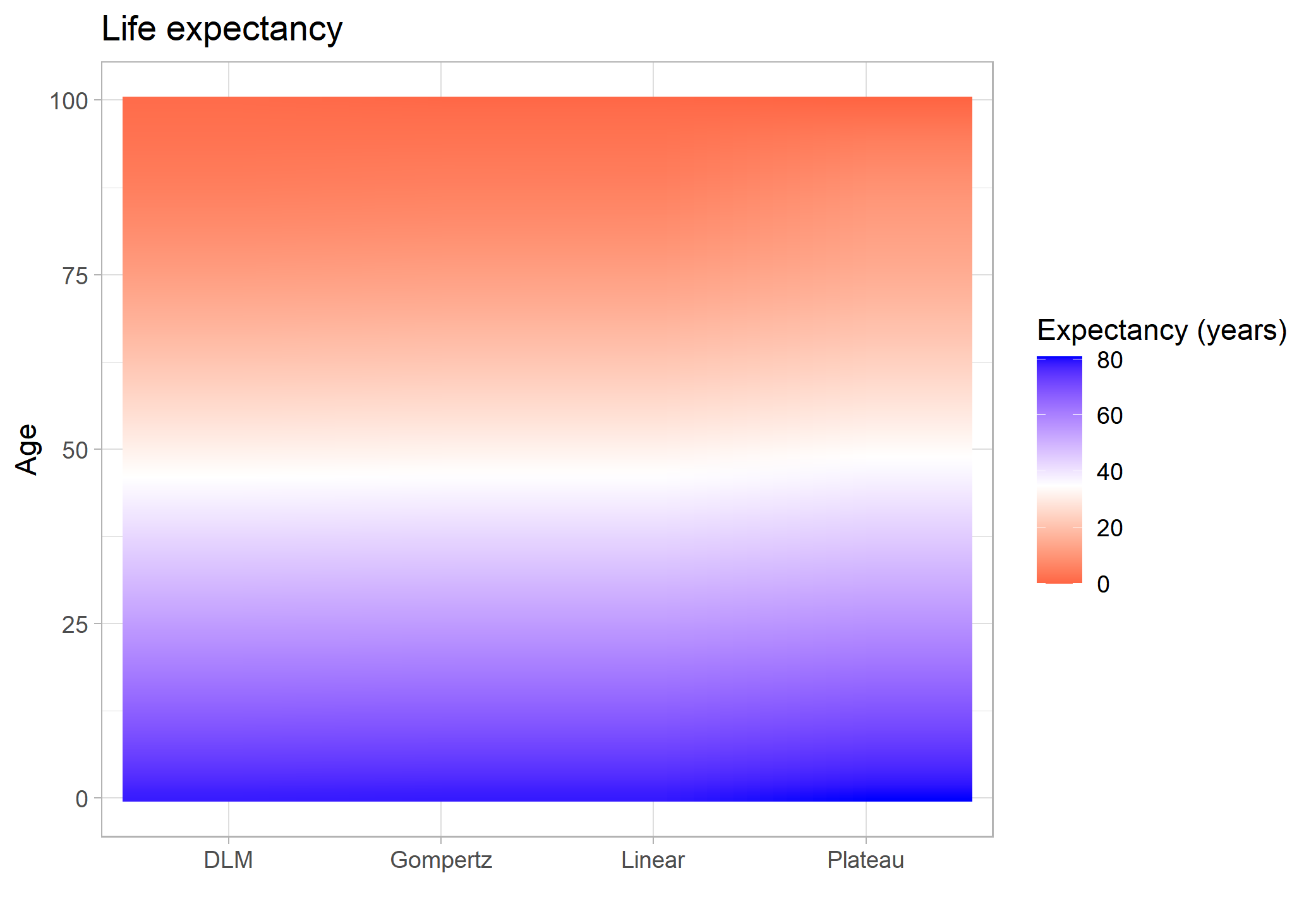}
      \caption{Posterior summaries via DLM: life expectancy for the United States. Total population, ages 0-100, the year 2019.}
    \label{fig:heatmap_closed_dlm}
\end{figure}

For the advanced ages modelling, the \code{"plateau"}, \code{"linear"} and \code{"gompertz"} methods are available for the \code{"DLM"} object through the \code{dlm_close} function. Usage and interaction with other functions is the same as seen in Section \ref{sec4.1} resulting in a \code{"ClosedDLM"} object, with the exception of \code{"new_Ex"} and \code{"new_Dx"} arguments that are not used in the DLM methods, as it models the log-mortality directly, replaced by the \code{"new_data"} argument. Consider the object \code{dlm_2019} and ages 80-100 for illustration:
\begin{CodeChunk}
\begin{CodeInput}
R> new_data <- log(new_Dx/new_Ex)
R> dlm.close1 <- dlm_close(dlm_2019, method = "plateau", max_age = 100)
R> dlm.close2 <- dlm_close(dlm_2019, method = "linear", max_age = 100,
                        new_data = new_data)
R> dlm.close3 <- dlm_close(dlm_2019, method = "gompertz", max_age = 100,
                        new_data = new_data)
\end{CodeInput}
\end{CodeChunk}
Here, we bring attention to the fact that, due to the model nature, it is possible to fit a reasonable advanced age curve without the closing methods. As long as the advanced age data available has some degree of reliability, we encourage the user to try fitting the simpler \code{dlm} function, as seen in Figure \ref{fig:closed_dlm}. Model comparison between different closing methods and their impact on life expectancy is shown in Figure \ref{fig:heatmap_closed_dlm}.
\begin{CodeChunk}
\begin{CodeInput}
R> new_Ex <- dplyr::filter(USA, Year == 2019)$Ex.Total[1:101]
R> new_Dx <- dplyr::filter(USA, Year == 2019)$Dx.Total[1:101]
R> new_y <- log(new_Dx/new_Ex)
R> dlm.fit3 <- dlm(new_y, delta = 0.85)
Simulating [===================================] 100% in  1m
R> plot(list(dlm.fit3, dlm.close1, dlm.close2, dlm.close3), plotIC = F,
     plotData = F, age = 70:100,
     labels = c("DLM fitted", "Plateau", "Linear", "Gompertz"),
     linetype = c("twodash","solid","solid","solid")) +
guides(colour = guide_legend(override.aes = list(linetype = c(6, 1, 1, 1))))
\end{CodeInput}
\end{CodeChunk}

\begin{CodeChunk}
\begin{CodeInput}
R> fits.new <- list(dlm.close1, dlm.close2, dlm.close3, dlm.fit3)
R> labels <- c("Plateau","Linear","Gompertz","DLM")
R> Heatmap(fits.new, x_lab = labels, age = 0:100)
\end{CodeInput}
\end{CodeChunk}

{\it Extrapolation for dynamic linear smoothers}

We consider  $k$-steps-ahead predictive distributions to extrapolate the fitted mortality curve by a Dynamic Linear Model. According to \citet[][Section 2.8]{petris2009dynamic}, for the DLM, the $k$-steps-ahead predictive distributions, $k=1,2,...$ are obtained as a by-product of the Kalman filter as follows:
\begin{eqnarray}\nonumber
  \pi(\theta_{x+k}|y_{1:x}) &=& \int \pi(\theta_{x+k}|\theta_{x+k-1}) \pi(\theta_{x+k-1}|y_{1:x}) d\theta_{x+k-1}  \\ \nonumber
  \pi(y_{x+k}|y_{1:x}) &=& \int \pi(y_{x+k}|\theta_{x+k}) \pi(\theta_{x+k}|y_{1:x}) d\theta_{x+k} \nonumber
\end{eqnarray}
where $\pi(.)$ are the filtered densities, $\pi(\theta_{x+k}|y_{1:x})$ denotes the k-steps-ahead prior distribution of the state $\theta$ and $\pi(y_{x+k}|y_{1:x})$ the k-steps-ahead forecast distribution of the observation. Considering  extrapolation  for $k$ ages ahead, we obtain the $h$-step-ahead prediction distributions, $h=1,2,\ldots,k$, conditional on information up to the maximum age used in the model fitting. The prediction can be obtained by the basic \code{predict()} function provided by \proglang{R} software as follows:
\begin{Code}
predict(object, h, prob = 0.95)
\end{Code}
This function receives an object of the class \code{"DLM"} adjusted by the function \code{dlm} and returns a \code{data.frame} with the death probability prediction and credible intervals, with credibility level specified by the argument \code{prob}  for the ages in the prediction horizon (argument \code{h}). Consider the 80-100 ages prediction for the \code{dlm_2019} object:
\begin{CodeChunk}
\begin{CodeInput}
R> dlm.fit4 <- predict(dlm_2019, h = 20, prob = 0.95)
R> head(dlm.fit4)
Ages  qx_fitted     qx_inf     qx_sup
1   81 0.04354295 0.03673579 0.05135104
2   82 0.05224183 0.04410836 0.06192136
3   83 0.05686888 0.04768670 0.06780340
4   84 0.06185414 0.05081621 0.07469514
5   85 0.06737384 0.05491058 0.08260258
6   86 0.07303701 0.05964790 0.09198139

R> plot(dlm_2019, plotIC = F, plotData = F) +
  geom_line(data = dlm.fit4,
            aes(x = Ages, y = qx_fitted, col = "Predict")) +
  geom_ribbon(data = dlm.fit4,
            aes(x = Ages, ymin = qx_inf, ymax = qx_sup, fill = "Predict"),
            alpha = 0.4) + 
  scale_color_manual(values = c("seagreen","red"),
                    label = c("DLM fitted", "Predict")) +
  guides(fill = "none") + labs(colour = "")
\end{CodeInput}
\end{CodeChunk}
\begin{figure}[!ht]
    \centering
    \includegraphics[width = 0.8\textwidth]{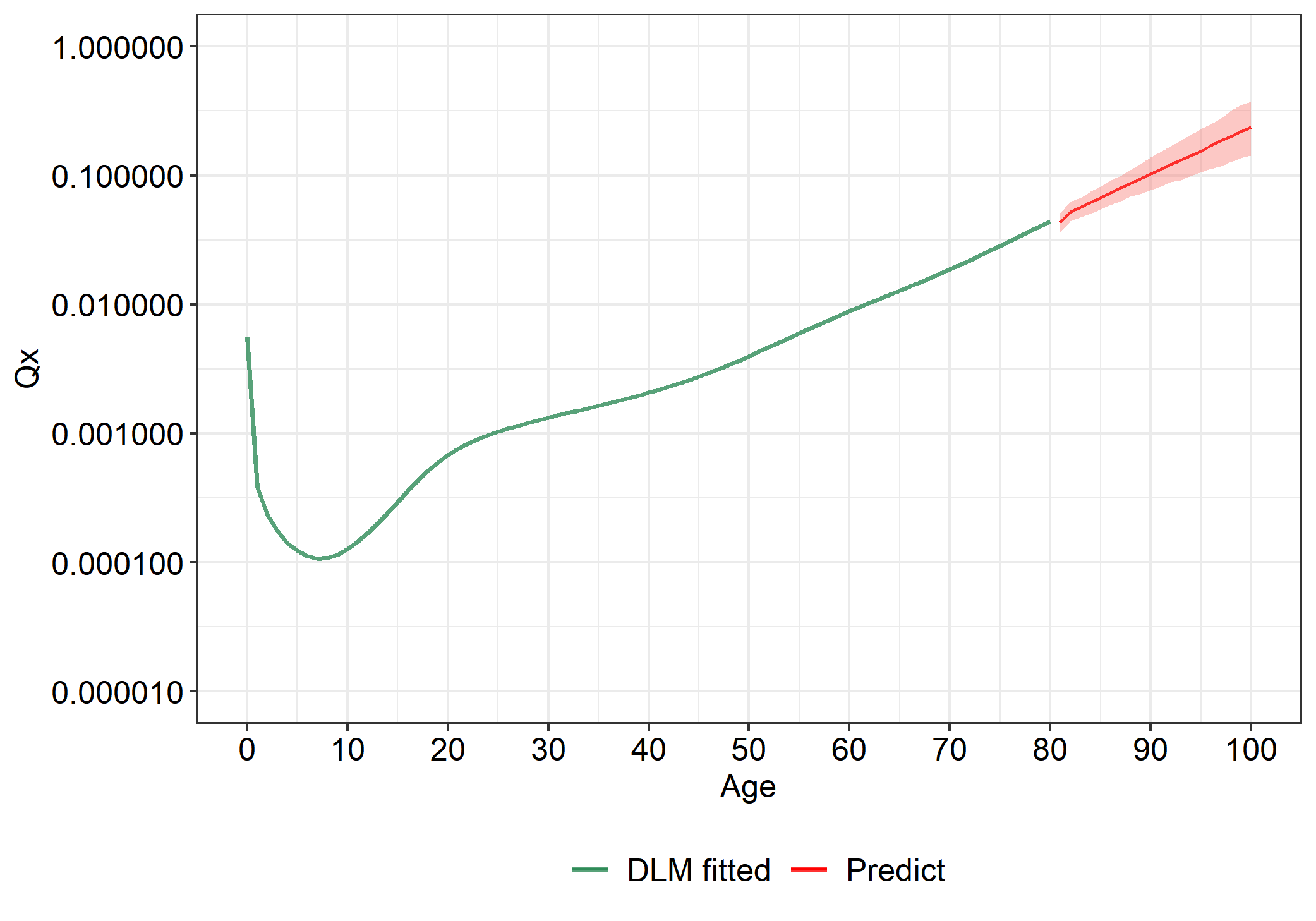}
      \caption{Posterior summaries via DLM: prediction for the United States. Total population, ages 0-100, the year 2019.}
    \label{fig:predict_dlm}
\end{figure}
The extrapolation via \code{predict} function is used in the life expectancy computation when the maximum age specified was not fitted. It replaces the extrapolation of the HP model method found in Section \ref{sec4.1} to match the \code{max_age} argument.

\section[Bayesian Lee-Carter model]{Bayesian Lee-Carter model}\label{sec5}

The methods presented in Sections \ref{sec2} and \ref{sec3} provide means for smoothing mortality rates over ages, but there are contexts in which mortality data are also available over years. Thus one can be interested in recognising the  evolution of mortality laws, as time passes.  Dynamic linear models, as described in Section \ref{sec3}, can be naturally used to accommodate time-indexed observations.  We consider a non-linear dynamic formulation indexed by age and time, as follows. Let $D_{x,t}$ denote the number of deaths at age $x$  and calendar period $t$; and $E_{x,t}$ denote the population exposed to risk at age $x$ and  time $t$. The basic Lee and Carter \citep{Lee1992} model seeks to describe the age-time surface of log mortality rates $y_{x,t}=\log\frac{D_{x,t}}{E_{x,t}}$ as:
\begin{equation}
{y}_{x,t}=\alpha_x+\beta_x \kappa_t+\varepsilon_{x,t}, \quad \varepsilon_{x,t} \stackrel{iid}{\sim} N(0,\sigma^2_{\varepsilon}),
\label{eqobs}
\end{equation}
\noindent where $\alpha_x$ denotes the general log-mortality pattern for age $x$; $\beta_x$ denotes an age-specific change rate in log-mortality; $\varepsilon_{x,t}$ are sequentially independent and homoscedastic random errors and $\kappa_t$ is an unobservable vector of time-indexed states, representing the global level of mortality at time period $t$, $t=1,2, \ldots; x=1,2,\ldots$. \cite{Pedroza2006} follows \cite{Lee1992} suggestion that $\kappa_t$, $t=1,2,\ldots$ evolve according to a random walk with drift, proposing the following state space representation for the temporal evolution of the states: 
\begin{equation}
\kappa_t=\theta+\kappa_{t-1}+\omega_t, \quad \omega_t \stackrel{iid}{\sim} N(0,\sigma^2_{\omega}),
\label{eqevol}
\end{equation}
where $\omega_t$ are independent and homoscedastic evolution random errors, which are independent of the observational errors $\varepsilon_{x,t}$. 

Unlike a traditional regression model, all quantities on the right side of (\ref{eqobs}) are unobservable. In order to ensure identifiability, \cite{Lee1992} impose the constraints $\sum_t\kappa_t=0$ and $\sum_x\beta_x=1$ and the estimation process uses single value decomposition to find a least squares solution, with $\kappa_t$ reestimated using Box-Jenkins methodology.  We follow \cite{Pedroza2006} in its fully Bayesian approach for the fit of the model given by (\ref{eqobs}) and (\ref{eqevol}), which enables simultaneous estimation of all the parameters while accounting for the uncertainty in the estimation process. Details on the MCMC algorithm can be found at \cite{Pedroza2006}.

\subsection[Dynamical graduation with BayesMortalityPlus]{Dynamical graduation with BayesMortalityPlus}

The \pkg{BayesMortalityPlus} package provides an \proglang{R} implementation of the Bayesian Lee Carter (BLC) model proposed by \cite{Pedroza2006}. The BLC models are constructed using the \code{blc()} function. The brief of this function is given by: 
\begin{Code}
blc(Y, prior = NULL, init = NULL, numit = 2000, warmup = 1000)
\end{Code}
The \code{blc} function prompts as input a \code{matrix} type dataset with the log mortality rates, where the columns represent years and lines represent ages, to create an object of the type \code{``BLC''} acting the Bayesian Lee-Carter model.
\begin{itemize}
    \item \code{Y} represents the matrix of log mortality rates containing the log ratio between deaths and exposures in a matrix format with ages on the rows and years on the columns.
    \item The argument \code{prior} (default= NULL) specifies the prior information about mean and variance, while \code{init} (default= NULL) specifies the initial values of each parameter to be estimated by the model. 
    \item The arguments \code{numit} (default=2000) and \code{warmup} (default=1000) control the number of iterations  and the warm-up  interval for the chains estimation, respectively.
\end{itemize}

\begin{figure}[t!]
    \centering
    \includegraphics[width = 0.9\textwidth]{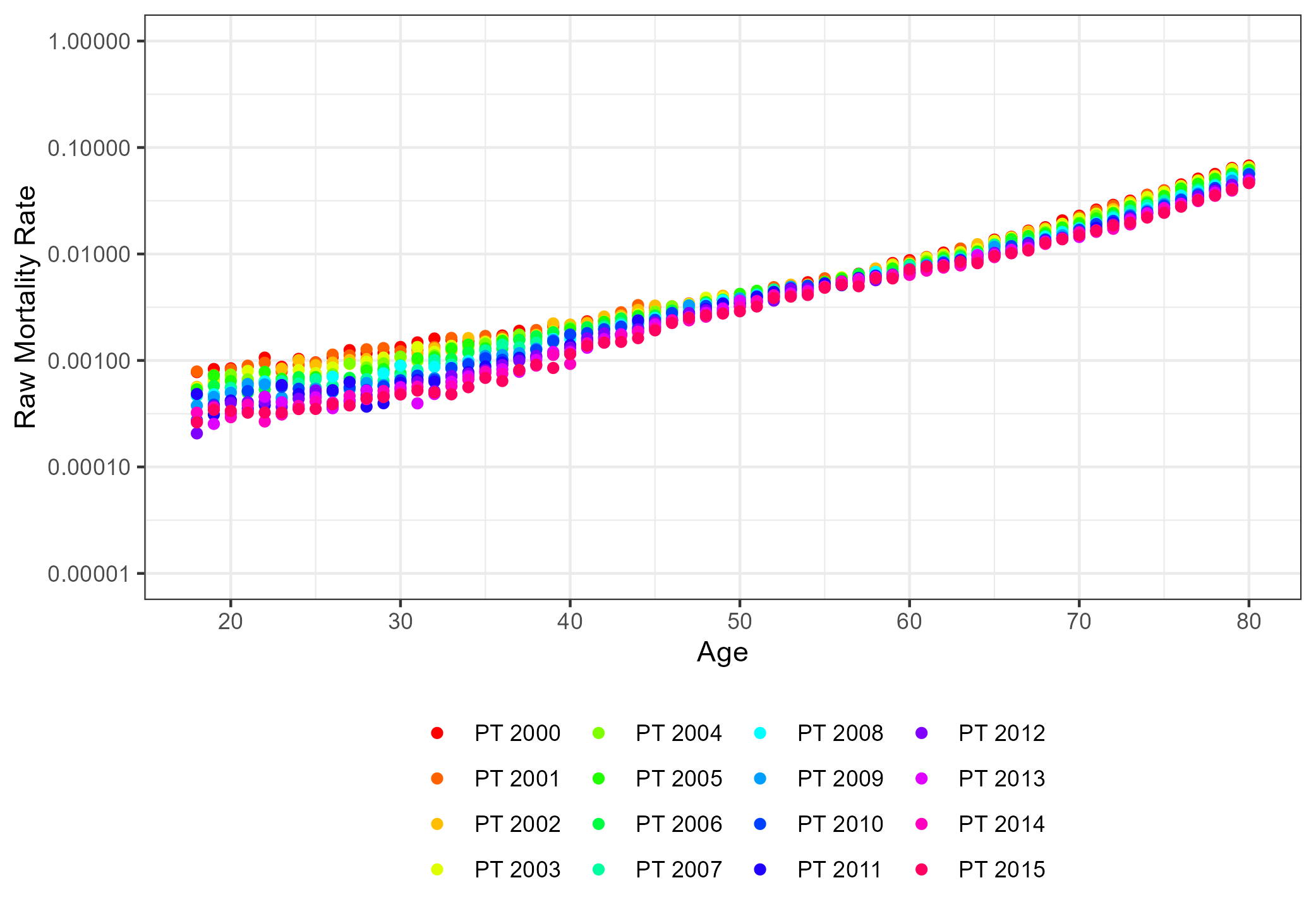}
    \caption{Raw mortality rates in log-scale. Portugal, total population, ages 18-80 and years 2000-2015.}
    \label{fig:raw_tables_pt}
\end{figure}

\begin{figure}[!hbt]
    \centering
    \includegraphics[width=0.9\textwidth]{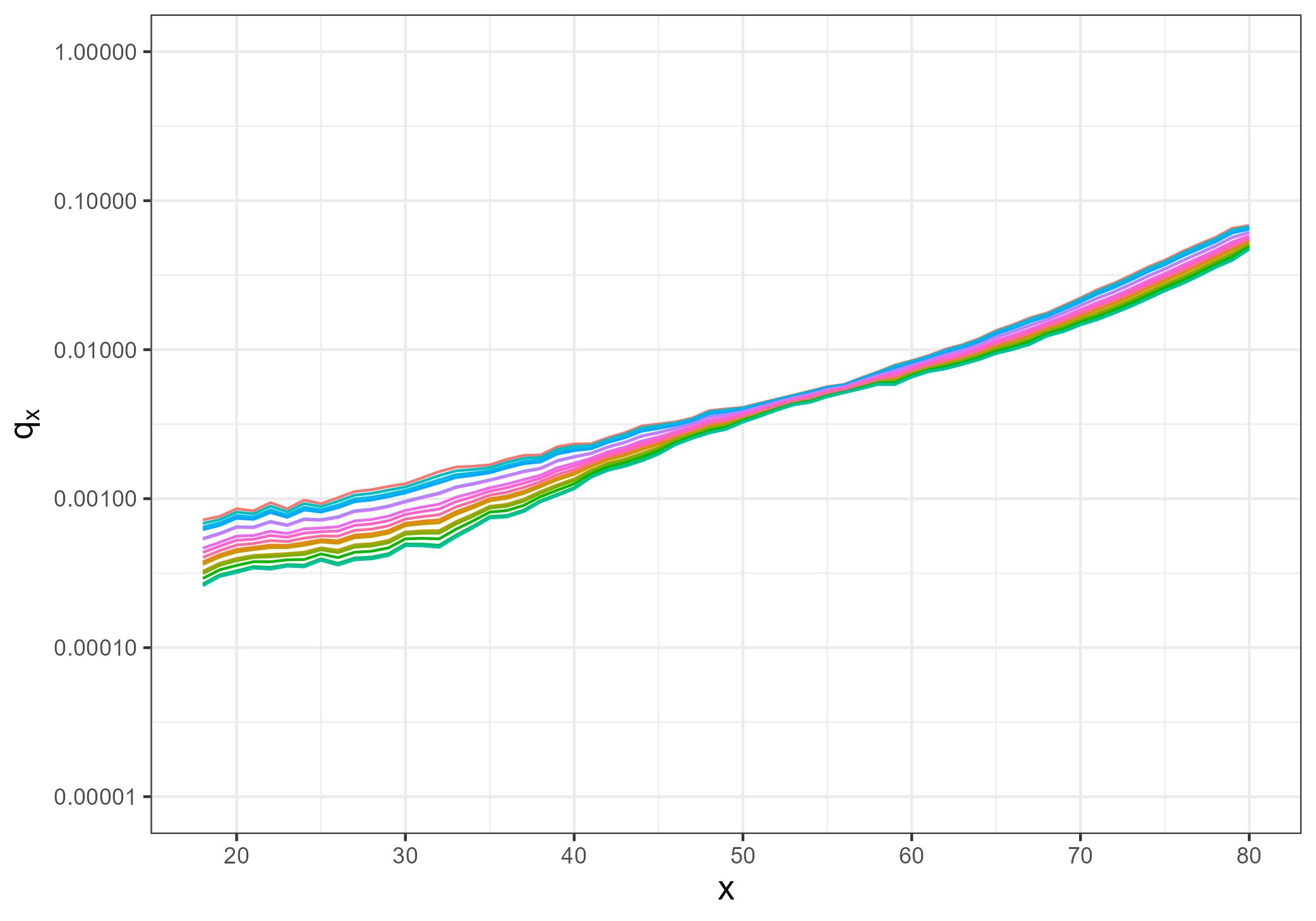}
    \caption{Posterior summaries via BLC: median mortality curves in log-scale. Portugal, total population, ages 18-80 and years 2000-2015.}
    \label{fig:blc_fitted}
\end{figure}

In order to illustrate the modelling of the Bayesian Lee-Carter and other features available on \pkg{BayesMortalityPlus} for mortality and life expectancy forecasting consider total mortality data from Portugal, for ages 18 to 80 and the period from 2000 to 2015, obtained from \cite{hmd}.  Figure \ref{fig:raw_tables_pt} illustrates the raw mortality rates over years via \pkg{ggplot} package. We consider the \pkg{tidyr} package to handle the data to plot. 
\begin{CodeChunk}
\begin{CodeInput}
R> data(PT)
R> Y <- PT
R> head(Y[,1:4])
        2000      2001      2002      2003
18 -7.159943 -7.145267 -7.480066 -7.498491
19 -7.092145 -7.216402 -7.331667 -7.380096
20 -7.077723 -7.101210 -7.244965 -7.287915
21 -7.035949 -7.020229 -7.370222 -7.133571
22 -6.845233 -6.956609 -7.363061 -7.184341
23 -7.043766 -7.150159 -7.083607 -7.302066
\end{CodeInput}
\end{CodeChunk}
\begin{CodeChunk}
\begin{CodeInput}
R> df.aux = tidyr::gather(data.frame(Y, idade = 18:80),
                       key = "Year", value = "log.qx", - idade)
R> ggplot(df.aux) +
    scale_y_continuous(trans = "log10", breaks = 10^-seq(0,5),
                     limits = 10^-c(5,0), labels = scales::comma) +
    scale_x_continuous(breaks = seq(0, 100, by = 10)) + 
    theme_bw() + theme(legend.position = "bottom") +
    labs(x = "Age", y = "Raw Mortality Rate", title = NULL) +
    geom_point(aes(x = idade, y = exp(log.qx), col = Year)) +
    scale_color_manual(name = NULL, values = c(rainbow(16)),
                     label = paste("PT", 2000:2015))
\end{CodeInput}
\end{CodeChunk}
The usage of the \code{blc} function to fit the Bayesian Lee-Carter model results in an object of class \code{"BLC"}. In this example, we consider the default settings to fit the BLC model with the code:
\begin{CodeChunk}
\begin{CodeInput}
R> fit.blc <- blc(Y, numit = 2000)
Simulating [===================================] 100% in 20s
\end{CodeInput}
\end{CodeChunk}
The \code{fitted} function returns the fitted log mortality estimates for each year. For the \code{"BLC"} object, the function returns an object of type \code{list} containing log mortality means (\code{\$mean}) as well as credible intervals (\code{\$lower} and \code{\$upper}). The \code{plot} function is called to visualize the evolution of the fitted log mortality estimates through the years (Figure \ref{fig:blc_fitted}). The output from the function \code{fitted} for ages 18-23 and years 2000-2003 and call to the \code{plot} function are shown as follows:
\begin{CodeChunk}
\begin{CodeInput}
R> head(fitted(fit.blc)[[1]][,1:4])
           2000         2001         2002         2003
18 0.0007230827 0.0006849759 0.0006474087 0.0006217643
19 0.0007610521 0.0007242883 0.0006878730 0.0006629022
20 0.0008571602 0.0008135041 0.0007703697 0.0007408658
21 0.0008284950 0.0007906396 0.0007529585 0.0007271215
22 0.0009426982 0.0008926108 0.0008432942 0.0008096202
23 0.0008562065 0.0008166684 0.0007775165 0.0007504693
R> plot(fit.blc, parameter = "fitted", ages = 18:80)
\end{CodeInput}
\end{CodeChunk}

\begin{figure}[!hbt]
    \centering
    \includegraphics[width=0.9\textwidth]{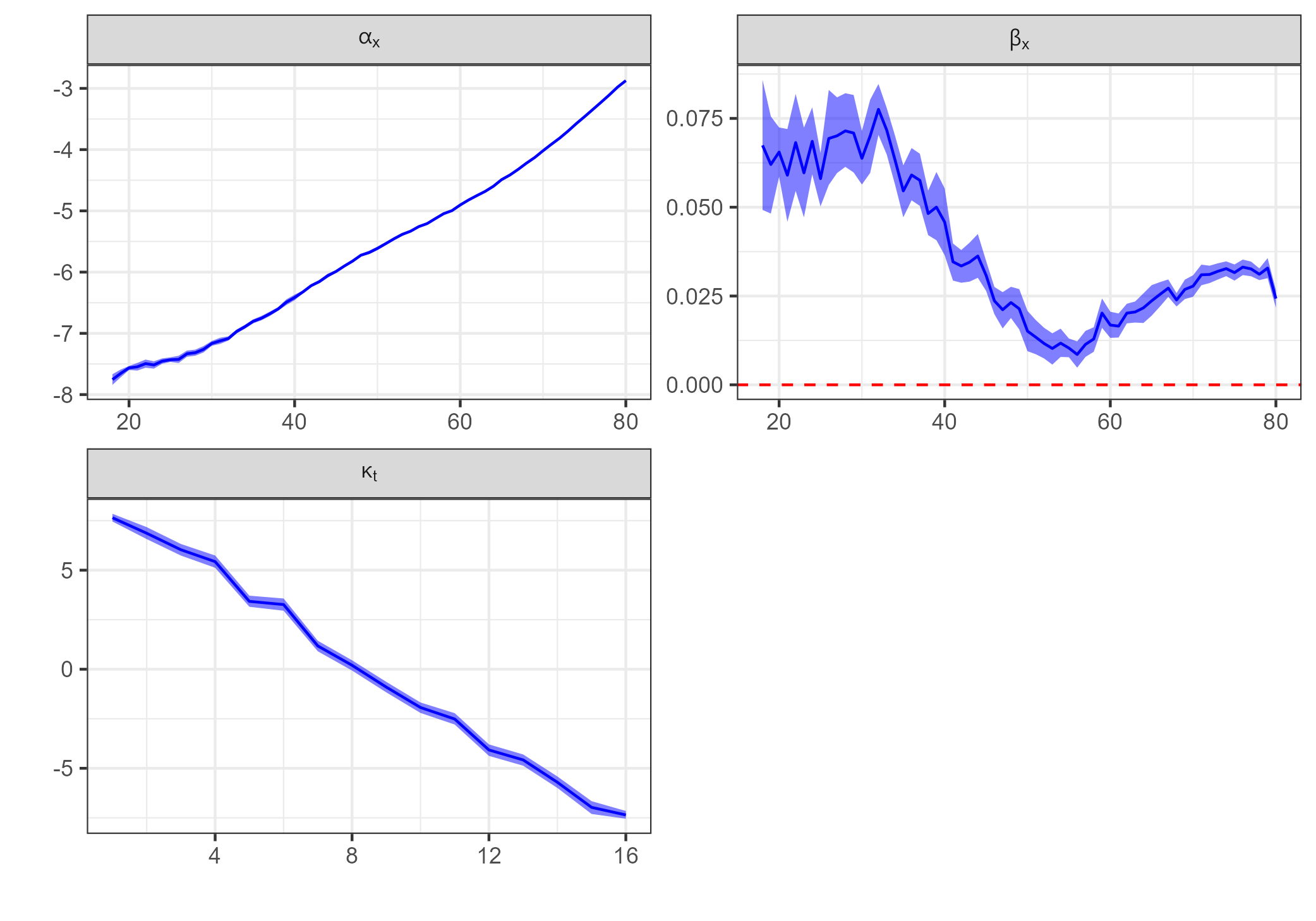}
    \caption{Posterior summaries via BLC: means and variance for each parameter of the fitted BLC model. Portugal, total population, ages 18-80 and years 2000-2015.}
    \label{fig:parameters_blc}
\end{figure}

The evolution of the mortality graduation can be seen through the  $\alpha$, $\beta$ and $\kappa$ parameters. Figure \ref{fig:parameters_blc} depicts the fitted parameters of the BLC model using the \code{plot} method for the \code{"BLC"} class.
\begin{CodeChunk}
\begin{CodeInput}
R> plot(fit.blc, parameter = "all", ages = 18:80)
\end{CodeInput}
\end{CodeChunk}

\begin{figure}[!hbt]
    \centering
    \includegraphics[width=0.9\textwidth]{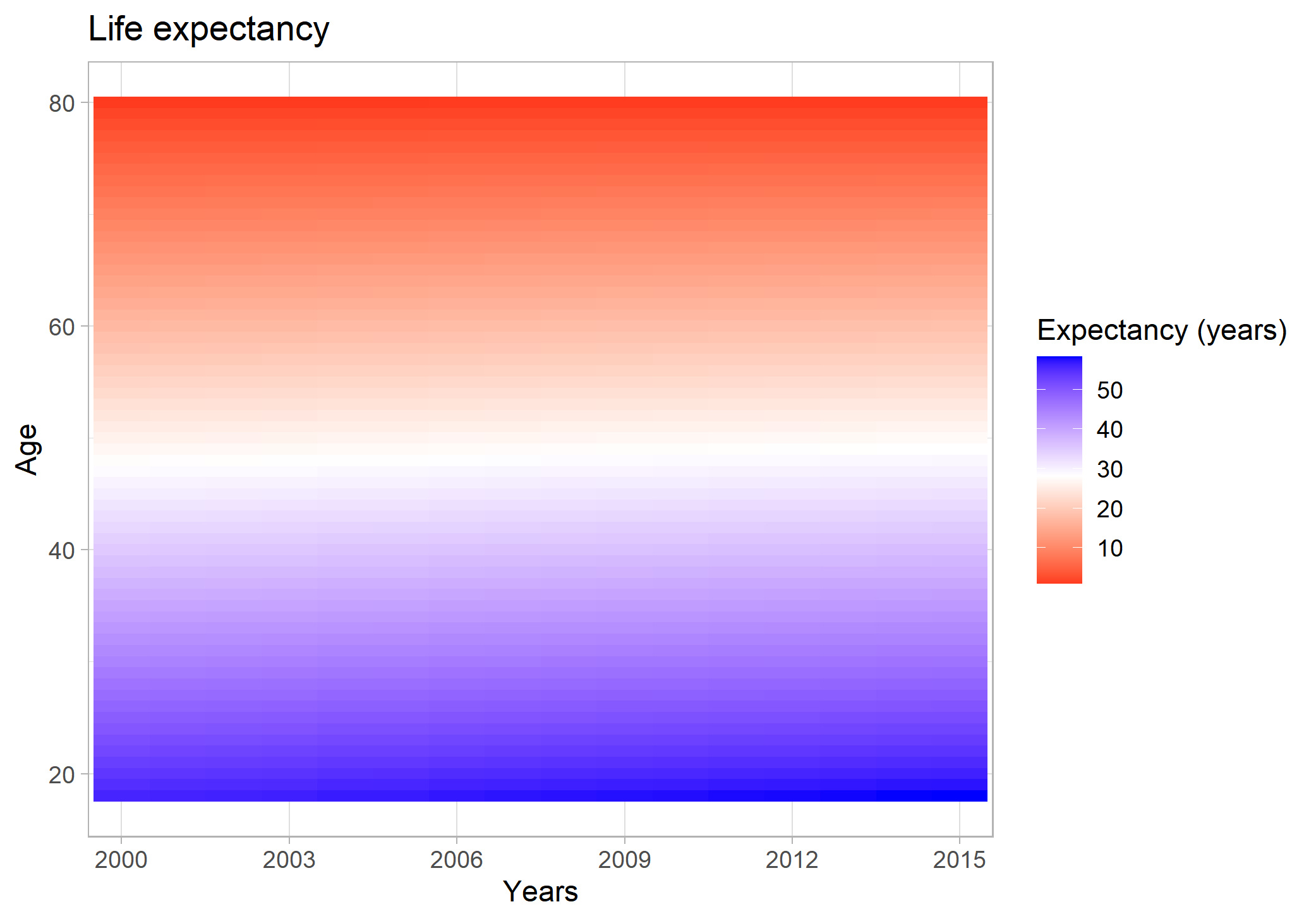}
    \caption{Posterior summaries via BLC: the life expectancy for Portugal country. Total population, 18-80 ages for years 2000-2015.}
    \label{fig:heatmap_blc}
\end{figure}

From Figure \ref{fig:parameters_blc} we see that the $\alpha_x$ parameter displays the general mortality pattern present in the data. Notice that the parameter $\kappa_t$ which represents the global level of mortality in period $t$ is decreasing through the years. This behaviour impacts the interpretation of parameter $\beta_x$ directly. In this case, $\beta_x$ is also called \textbf{improvement} and reflects the rate at which mortality is decreasing over the years. In the package \pkg{BayesMortalityPlus} the improvement of the BLC model is implemented via the function \code{improvement}. This function estimates the improvement percentage for each age throughout the whole considered time interval. 
\begin{CodeChunk}
\begin{CodeInput}
R> head(improvement(fit.blc, cred = 0.95))
  improvement  lower.lim  upper.lim
1  0.06566689 0.04441260 0.08594036
2  0.06024696 0.04458217 0.07688504
3  0.06355622 0.05544481 0.07100160
4  0.05708944 0.04320909 0.07150397
5  0.06620879 0.05008862 0.08211478
6  0.05758188 0.04366629 0.07128359
\end{CodeInput}
\end{CodeChunk}
The \code{expectancy} and \code{Heatmap} methods are available to compute life expectations and their uncertainty via credible intervals for each age and year. For instance, we consider specific ages using the argument \code{``at''} and the output can be obtained with the commands:
\begin{CodeChunk}
\begin{CodeInput}
R> expectancy(fit.blc, at = c(1,21,41,61))expectancy[,1:4]
     2000   2001   2002   2003
18 55.844 55.997 56.153 56.262
38 36.942 37.042 37.145 37.218
58 18.983 19.053 19.124 19.175
78  2.646  2.655  2.663  2.669
\end{CodeInput}
\end{CodeChunk}
The plot of life expectancy (see Figure \ref{fig:heatmap_blc}) can be produced using the code: 
\begin{CodeChunk}
\begin{CodeInput}
R> Heatmap(fit.blc, x_lab = 2000:2015, age = 18:80) + 
  scale_x_discrete(breaks = seq(2000,2015, by=3))
\end{CodeInput}
\end{CodeChunk}

\vspace{0.5cm}
\noindent {\it Forecast for fitted BLC models} 

In the package \pkg{BayesMortalityPlus} the forecasting of the BLC mortality model for n-years ahead is implemented via the \code{predict} function. Following \cite{Pedroza2006} the predictive steps can be incorporated into the Gibbs sampler. The posterior predictive distribution $(y_{n+1}\mid Y^n)$ for future observations can be expressed as
$$
p(y_{n+1}\mid Y^n) = \int p(y_{n+1}\mid \phi, Y^n)p(\phi \mid Y^n)d\phi = \int p(y_{n+1}\mid \phi)p(\phi \mid Y^n)d\phi
$$
where $\phi$ represents the model parameters. We assume that $y_{n+1}$ and $Y^n$ are conditionally independent given $\phi$.

The prediction can be obtained by the basic \code{predict} function, specifying the years ahead to be forecasted by the \code{h} argument, resulting in an object of class \code{"PredBLC"}. The output considers 10-years-ahead ($h=10$) for the Portugal mortality experience: 
\begin{CodeChunk}
\begin{CodeInput}
R> fit.blc2 <- predict(fit.blc, h = 10)
R> print(fit.blc2)
Forecast of a Bayesian Lee-Carter model (h = 10)
\end{CodeInput}
\end{CodeChunk}

The functions such as \code{fitted}, \code{expectancy} and \code{Heatmap} also are available for the \code{"PredBLC"} object. See below:
\begin{CodeChunk}
\begin{CodeInput}
R> head(fitted(fit.blc2)$mean[,1:4])
             [,1]         [,2]         [,3]         [,4]
[1,] 0.0002432606 0.0002246182 0.0002126753 0.0001973058
[2,] 0.0002832726 0.0002652922 0.0002474990 0.0002300914
[3,] 0.0003001941 0.0002814294 0.0002644005 0.0002453596
[4,] 0.0003242555 0.0003062827 0.0002880610 0.0002717220
[5,] 0.0003149540 0.0002957785 0.0002739343 0.0002563254
[6,] 0.0003343524 0.0003122069 0.0002932249 0.0002785498

R> expectancy(fit.blc2, at = c(1,21,41,61))$expectancy[,1:4]
       [,1]   [,2]   [,3]   [,4]
[1,] 58.274 58.389 58.501 58.615
[2,] 38.679 38.770 38.858 38.950
[3,] 20.200 20.265 20.327 20.392
[4,]  2.779  2.786  2.792  2.798

R> Heatmap(fit.blc2, x_lab = 2016:2025, age = 18:80) + 
  scale_x_discrete(breaks = seq(2016,2025, by=3))
\end{CodeInput}
\end{CodeChunk}
\begin{figure}[!hbt]
    \centering
    \includegraphics[width=0.9\textwidth]{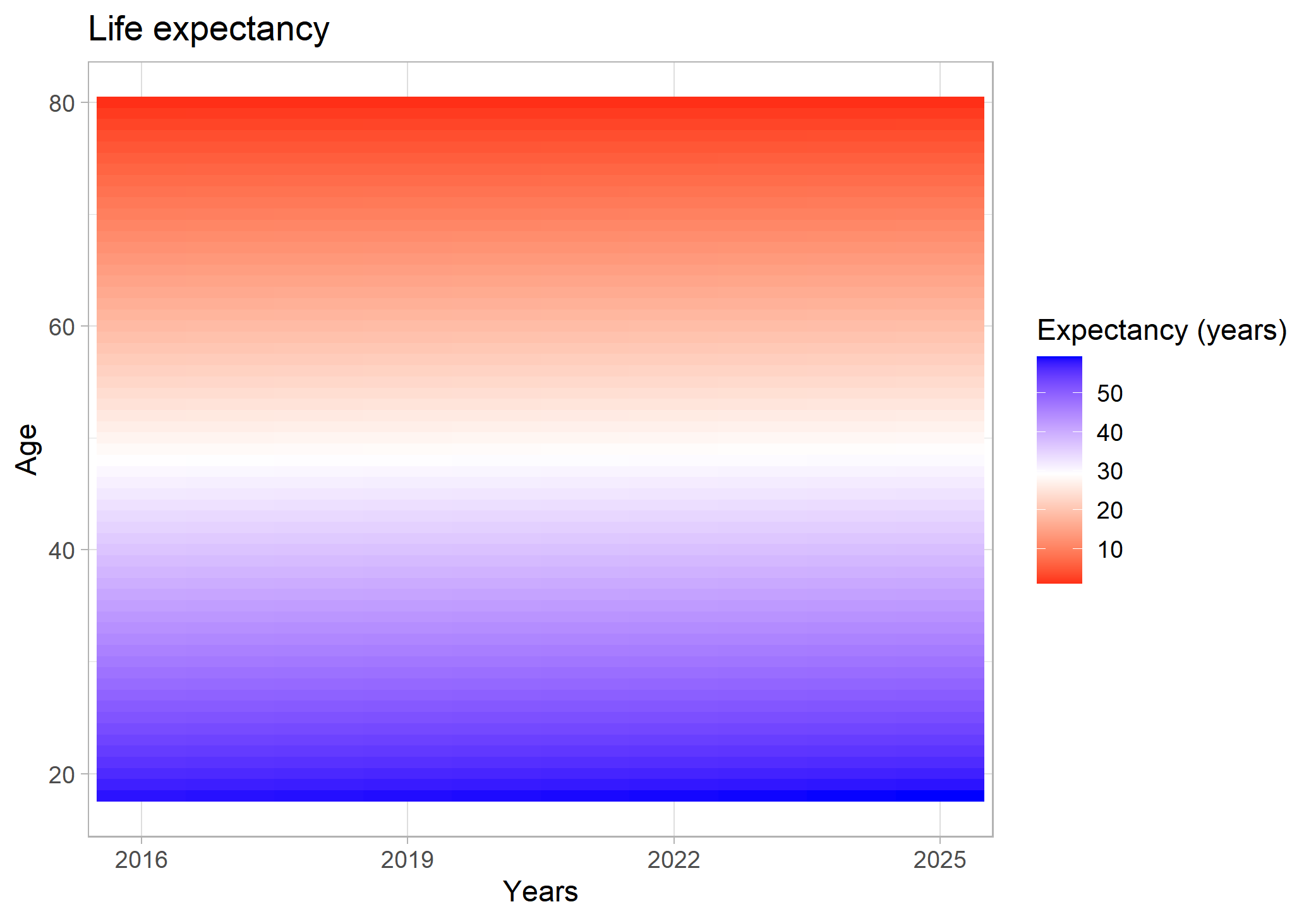}
    \caption{Posterior summaries via BLC: Forecasted life expectancy for Portugal country. Total population, 18-80 ages for years 2016-2025.}
    \label{fig:heatmap_pred_blc}
\end{figure}

{\color{blue}}

\section[Conclusions]{Conclusions}\label{sec6}
In this paper, we present an \proglang{R} package called \pkg{BayesMortalityPlus} for mortality modelling using a Bayesian approach. The package allows for Bayesian inference for several models used for mortality table graduation as well as mortality prediction for future years. The tools available in the proposed package provide model fitting, visualization of parameters and linear and non-linear functions of parameters, and uncertainty quantification via credible intervals or complete posterior distributions. Examples are provided to illustrate the features of all models. For the Heligman-Pollard law of mortality, Bayesian model fitting is available for three probability distributions: Poisson, Binomial and Log-Normal. The interpretable parameters in the HP model can be visualised through graphs and summaries of the resulting posterior distributions. As an alternative to spline fitting for mortality graduation, the \code{dlm} function takes into account the autocorrelation in the mortality across ages and  provides estimation of mortality curves and extrapolation for older ages. Opposed to the HP model, DLM fit does not depend on a specific law of mortality for the data. Smoothness is controlled by discount factors, which are common practice in the context of time series modelling via dynamic models, and offer flexibility to the model of death probabilities. Lastly, the Bayesian version of the well-known Lee-Carter model is implemented via MCMC methods and the \code{predict} function allows for prediction in future time steps. Furthermore, point estimates, as well as uncertainty measurements, can be computed for improvement parameters which are often the main interest in studies of longevity and pricing of products of long term in insurance modelling.

\section*{Acknowledgments}

We are grateful to LabMA/UFRJ (Laborat\'orio de Matem\'atica Aplicada of the Universidade Federal do Rio de Janeiro) based in Brazil for financial support and to its members for the very enriching discussions.

\bibliography{references}

\end{document}